\documentclass[aps,pre,twocolumn,superscriptaddress,showpacs]{revtex4-1}

\usepackage{graphicx}
\usepackage{subfigure}

\begin{document}


\title{Assessing the consistency of community structure in complex networks }



\author{Matthew Steen}
\affiliation{Department of Radiology, Wake Forest University School of 
Medicine, Winston-Salem, North Carolina, USA}

\author{Satoru Hayasaka}
\affiliation{Department of Biostatistical Sciences, Wake Forest University 
School of Medicine, Winston-Salem, North Carolina, USA}

\author{Karen Joyce}
\affiliation{School of Biomedical Engineering and Sciences, Wake Forest 
University School of Medicine, Winston-Salem, North Carolina, USA} 

\author{Paul Laurienti}
\affiliation{Department of Radiology, Wake Forest University School of 
Medicine, Winston-Salem, North Carolina, USA}


\date{\today}

\begin{abstract}
In recent years, community structure has emerged as a key component of complex 
network analysis. As more data has been collected, researchers have begun 
investigating changing community structure across multiple networks. Several 
methods exist to analyze changing communities, but most of these are limited to 
evolution of a single network over time. In addition, most of the existing 
methods are more concerned with change at the community level than at the level 
of the individual node. In this paper, we introduce scaled inclusivity, which 
is a method to quantify the change in community structure across networks. 
Scaled inclusivity evaluates the consistency of the classification of 
every node in a network independently. In addition, the method can be applied 
cross-sectionally as well as longitudinally. In this paper, we calculate the 
scaled inclusivity for a set of simulated networks of United States cities and 
a set of real networks consisting of teams that play in the top division of 
American college football. We found that scaled inclusivity yields reasonable 
results for the consistency of individual nodes in both sets of networks. We 
propose that scaled inclusivity may provide a useful way to quantify the change 
in a network's community structure.

\end{abstract}


\maketitle


\section{Introduction}
In recent years, the study of real-world complex networks has increased 
dramatically; social networks \cite{social}, the world wide web \cite{web}, and 
faculty collaboration networks \cite{collaboration} are among the most commonly 
studied.

An offshoot of this expansion has been the study of community structure in 
complex networks. Introduced in 2002 by Girvan and Newman \cite{modularity}, 
the community structure of a network helps make sense of the 
interactions between the nodes. In that paper, the authors analyze a network 
consisting of the members of a karate club that split into two factions, a 
faculty collaboration network across Europe, and several others. Mathematical 
community analysis was able to identify the two factions of the karate club and 
split the collaboration network into several geographically coherent groups.

\begin{figure}[h]
\begin{center}
\subfigure[Network 1]{\label{fig:edge-a}\includegraphics[width=40mm,height=30mm]
{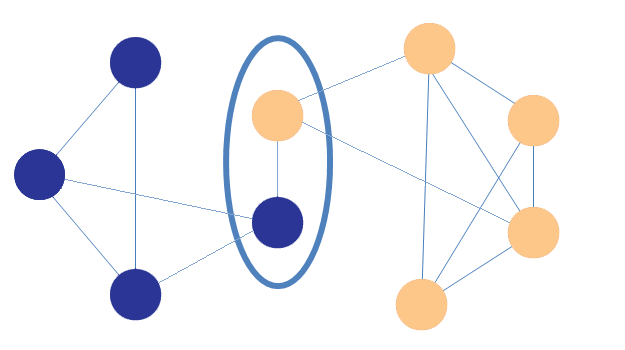}}
\subfigure[Network 2]{\label{fig:edge-b}\includegraphics[width=40mm,height=30mm]
{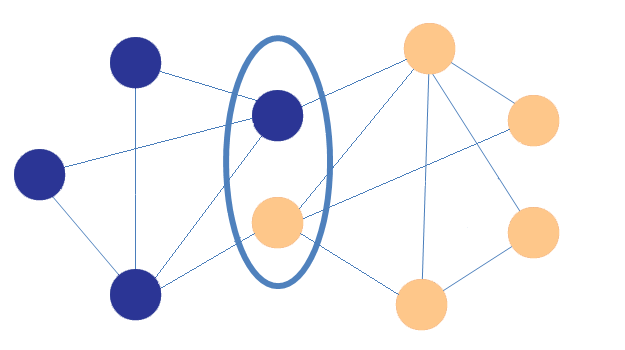}} \\
\end{center}
\caption{(Color online)  Consider the two networks in (a) and (b). Both are 
realizations of a network with the same underlying nodes. Several edges change 
between the two, and two nodes switch communities (identified by node color). 
In a large network and across many realizations, it is easy to imagine such 
minor shifts occurring many times. The problem at hand is how to quantify the 
consistency of the communities identified across several such realizations.
\label{net}}
\end{figure}

In the past few years, more complex datasets have been studied, and the 
change in community structures of networks over time has become an area of 
focus. Fig. \ref{net} shows an example of changing community structure in two 
realizations of a simple network. Such shifts are easy to imagine on a larger 
scale in a complex network with many realizations. 

The consistency (or lack thereof) of communities in a given network over time 
can be a good indicator of large-scale change in a network. Hopcroft, et al.
\cite{Hopcroft} studied the changes in a network consisting of journal articles 
from the NEC CiteSeer database and their references. The communities 
correlated well with certain fields of study, and the emergence of new 
communities over time was shown to mirror the emergence of new fields in the 
literature.

While change in the community structure of a network has recently become an 
object of study, many of the methods proposed explicitly or implicitly make 
assumptions based on the idea of change over time \cite{Palla, Chakrabarti, 
Asur, Fenn}. These methods do not lend 
themselves well to the analysis of multiple realizations of a single network. 
The assumption that networks progress from one to the next in a linear fashion 
is not appropriate when the networks are not linearly related.

In addition, the existing methods spend relatively little time on how best to 
identify the same community across several networks; in some cases, it is 
assumed that communities are sufficiently consistent over time to make this a 
moot point. Across networks that are not linearly related, this assumption 
does not hold. The variation in the community structure of multiple networks 
can be significant, so a more rigorous way to quantify the consistency of 
community structure is necessary.

In this paper, we propose a method for comparing the consistency of community 
structure 
across different realizations of a network, be they the same network over time 
or simultaneous realizations of a single network. In particular, we propose a 
method to describe how consistently each node is part of the same community 
across different partitions, which we call scaled inclusivity. This method 
enables us to identify which nodes tend to remain in the same community in 
different network partitions, forming a "core" of that community. Likewise, the 
method also allows identification of transient nodes that become part of 
different communities across partitions.

The remainder of this manuscript is organized as follows: Section II describes 
the analysis algorithm, Section III describes the simulated and real data 
networks used, Section IV describes the results of the algorithm applied to 
both simulated and real data, and Section V discusses the implications of this 
work. 


\section{Methods}
Community structure analysis is a difficult problem. There are many community 
detection methodologies, and finding better algorithms is an area of ongoing 
research. We use the QCut algorithm, but any community 
detection algorithm could be used. We only evaluate a method that places each 
node into exactly one community, and the method we propose here requires that 
each node be in at least one community. 
The most common metric used to evaluate community structure, 
modularity, was introduced by Newman and Girvan \cite{Q}: \begin{equation} 
Q=\sum_{i=1}^{m} \left[ \frac{e_{ii}}{M} - \left( \frac{a_i}{M} \right)^2 
\right]\label{Q} \end{equation} where $m$ is the number of modules in the 
network, 
$e_{ii}$ is the total number of intra-modular edges in module $i$ ($i=1,2,
\cdots,m$) , $a_i$ is the total degree for the vertices in $i$, and $M$ is 
twice the total number of edges (the sum of the degree for all vertices in the 
network). Maximizing $Q$ has been proven to be an NP-hard problem \cite{NP}, 
meaning that the only way to guarantee an optimal solution is to try all 
possible solutions. Because this is impractical for all but the smallest 
networks, various algorithms have been created to balance optimization of $Q$ 
with run time. In our analysis, we used the QCut algorithm introduced in 
\cite{Ruan}. It should be emphasized that we are not attempting to evaluate or 
demonstrate the effectiveness of the QCut algorithm, but rather use QCut as a 
method to identify community structure. 

The method described here is used to assess consistency of partitions 
on a set of networks assumed to have a similar underlying structure or on a 
series of networks with alterations in the structure over time. While the 
former is a collection of multiple realizations of the networks without any 
ordering (e.g., metabolic networks from different samples or brain connectivity 
networks from multiple subjects), the latter has a particular ordering of the 
networks in which one network is a rewired version of the prior one. In both 
cases, it is assumed that the nodes are constant and that there exists 
a true partition of the nodes which may or may not be explicitly known.

\subsection{Identify best partition}

The first step is to identify the best partition for each of $n$ realizations 
of the network with $v$ nodes. The goal of this step is to find a reasonable 
partition of the network since each run of the QCut algorithm (like many 
community structure algorithms currently available) may produce slightly 
different partitions. This step starts by generating multiple  
partitions by the QCut algorithm --- $g$ times on a network $G_i$ ($i=1,2, 
\cdots,n$). Because the goal of any modularity-based algorithm is to maximize 
Q (Eq. \ref{Q}), the run that produced the partition with the highest Q value 
for the network is chosen as the best partition for this network, denoted by 
$Q'_i$. It should be noted that this step is unnecessary for deterministic 
community structure algorithms.

For nondeterministic community structure algorithms that are not based on 
optimization of a single parameter, an alternative method is to choose the 
partition that is most similar to the others. We propose the Jaccard similarity 
index to determine similarity. 

Let $Q_i^j$ be the $j$-th partition ($j=1,2,\cdots,g$) of network $i$. 
For any pair of partitions $Q_i^c$ and $Q_i^d$ ($c\neq d\in 1,2,\cdots,g$), the 
similarity of the  partitions is assessed by calculating the 
Jaccard similarity index \begin{equation} J_{cd} = \frac{|S_c \cap S_d|}{|S_c 
\cup S_d|} \label{Jaccard} \end{equation} where $S_c$ and $S_d$ are sets of 
node pairs with the same community memberships in partitions $Q_i^c$ and 
$Q_i^d$, 
respectively. For example, in partition $Q_i^c$, if nodes $q$ and $r$ are 
in the same community (i.e., $m_q=m_r$), then that node pair $(q,r)$ is 
included in the set $S_c$. The advantage of the Jaccard index is that it does 
not depend on the arbitrary numbering of communities across partitions. Even if 
community 5 in $Q_i^c$ may correspond to community 11 in $Q_i^d$, the Jaccard 
index can assess the similarity between partitions without re-assigning 
community numbers. Calculating $J_{cd}$ for all pairs of partition results in 
matrix $J$ whose $c,d$-th element is $J_{cd}$ if $c\neq d$ and 0 if $c=d$.

\subsection{Assess consistency of classification}
The next step assesses the consistency of a single partition when compared with 
the rest of the group. That single partition is chosen from the group and is 
denoted by $Q_R$; it is compared to $Q'_i$ for $i=1,2,\cdots,n$, where $Q'_i 
\neq Q_R$. To do so, consistency between communities in $Q'_i$ and $Q_R$ is 
first assessed at the community level. Assume $Q'_i$ consists of $y$ 
communities and $Q_R$ consists of $z$ communities. Then a 
community-by-community similarity matrix $X_{iR}$ is calculated, with its 
$p,q$-th element ($p=1,2,\cdots,y$), ($q=1,2,\cdots,z$) calculated as 
\begin{equation} X_{iR}^{pq}=\frac{|A_p \cap R_q|}{|A_p|} 
\frac{|A_p \cap R_q|}{|R_q|} \end{equation} where $A_p$ is the set of nodes 
belonging to the $p$-th community in $Q'_i$ (i.e., $\left\{ a | m_a=p 
\right\}$) and $R_q$ is the set of nodes belonging to the $q$-th community in 
$Q_R$ (i.e.,$\left\{ b | m_b=q \right\}$). The resulting values range from 
0 to 1, where 0 indicates zero overlap between the two communities and 1 
represents no change in the member nodes. This metric is preferable to the 
relative overlap (intersection over union) because it yields a lower value in 
the case of partial overlap between a large community and a smaller community. 
In other words, it more harshly penalizes poor specificity. The similarity 
matrix $X_{iR}$ is calculated for all realizations $i=1,2,\cdots,n$ where $Q'_i 
\neq Q_R$, and is used to summarize consistent community membership, relative 
to the reference partition, at the nodal level. A vector of length $v$ denoted 
as $V_m$, with each element corresponding to each node, is used to record the 
consistency in community memberships. We consider three different ways to 
assess this: binary exclusivity, binary inclusivity, and scaled inclusivity.

\begin{figure}[h]
\begin{center}
\subfigure[Network 1]{\label{fig:edge-a}\includegraphics[width=40mm,height=30mm]
{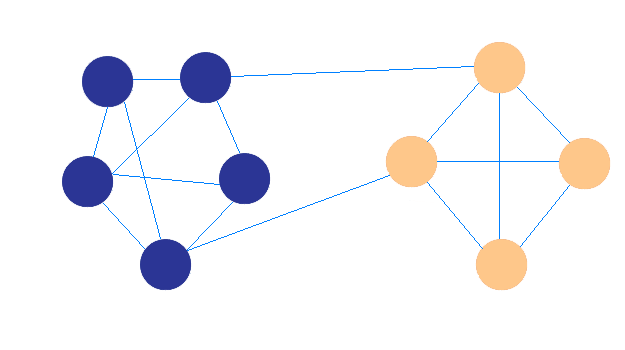}}
\subfigure[Network 2]{\label{fig:edge-b}\includegraphics[width=40mm,height=30mm]
{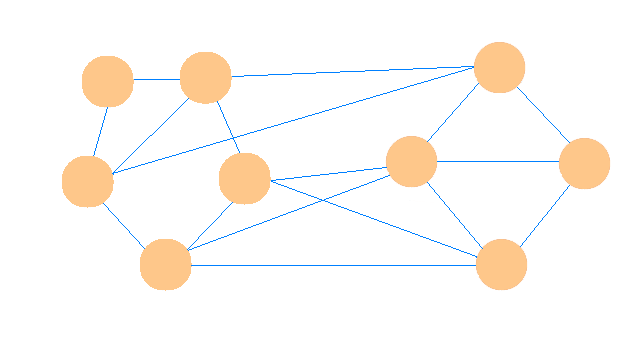}} \\
\end{center}
\caption{(Color online) Depicted are two communities in (a) merging into one 
community in (b) due to increased intercommunity edges. In binary exclusivity, 
the blue (dark gray) community would be counted as the better match since the 
number of 
nodes is greater. These nodes would be given a value of 1 in the scaled 
inclusivity map, and the orange (light gray) nodes would be given a value of 0. 
In binary 
inclusivity, both sets of nodes would be given a value of 1. In scaled 
inclusivity, assuming the second network is the reference, the blue (dark gray) 
nodes would 
all receive a value of $\frac{5}{5}\cdot\frac{5}{9} = 0.\overline{5}$, and the 
orange (light gray) nodes would receive a value of $\frac{4}{4}\cdot
\frac{4}{9} = 0.
\overline{4}$, thereby fairly scaling the scores of all nodes. This metric 
also allows for differentiation between merging communities as shown here and 
one community that does not change between two networks, which would have a 
value of 1 for all nodes. 
\label{merge}}
\end{figure}

In binary exclusive classification, a single community in each realization 
$Q'_i$ is identified as the best match for each community in the reference 
partition $Q_R$. This is done by finding the maximum in each column of the 
similarity matrix $X_{iR}$, corresponding to community $q$ in the reference 
partition. If the column maximum occurs on the $p$-th row, that indicates 
community $p$ of partition $Q'_i$ corresponds best to community $q$ of the 
reference partition $Q_R$. For all the nodes corresponding to community $p$ in 
partition $Q'_i$, a value of 1 is added to the corresponding elements of the 
recording vector $V_m$. This process is repeated for all communities in $Q_R$, 
and the occurrence of the best match is counted at the nodal level. The 
partitions from all realizations $Q'_j$, $j=1,2,\cdots,n$ are compared against 
$Q_R$, producing the final value of the recording vector $V_m$ with each nodal 
value summarizing the number of times it was "correctly" classified relative to 
the reference partition $Q_R$. Although this approach is intuitive, it fails to 
account for cases where multiple communities roughly split a community in the 
reference partition. For example, in time series data, it is possible that two 
communities will merge to form a larger community in the reference partition 
(see Fig. \ref{merge}); it seems inaccurate to only count the larger of these 
two communities as correctly classified.

An alternative approach is binary inclusivity, which is similar to binary 
exclusivity. However, two communities in one network that evenly split a 
community in the referent network can both be counted as correct 
classifications. As in binary exclusivity, the similarity matrix $X_{iR}$ is 
used to calculate the similarity of the communities, but all communities with 
some overlap are included. One could restrict this to communities with 
extensive overlap, but 
some threshold would have to be determined for the similarity value of the 
communities in question. For all the nodes belonging to the communities 
identified as the best matches, a value of 1 is added to the corresponding 
elements in the recording vector $V_m$. The process is repeated for all the 
communities in the reference partitions $Q_R$ for partitions from all the 
realizations $Q'_i$, as in binary exclusivity described above. One 
disadvantage of this method is that a very large community can include a part 
of a small community, but all common elements are considered equally correct. 
In addition, without a threshold, all nodes are assigned the same value at 
every step, which yields no useful information. Appropriately defining a 
threshold is a difficult and subjective task, and potentially small differences 
between some communities included and excluded are not reflected by assigning 
binary values.

Finally, scaled inclusivity takes into account any matching with any community 
in the reference partition. While binary exclusivity and inclusivity only add a 
value of 1 to nodes corresponding to the matching communities, scaled 
inclusivity adds the value $X_{iR}^{pq}$ to $A_p \cap R_q$ (i.e., where 
$X_{iR}^{pq}>0$). Thus, how well a node is classified in any given realization 
is scaled based on how well its communities in the two networks match. As in 
binary exclusivity and inclusivity, this process is repeated for all the 
communities in the reference partition $Q_R$ for partitions from all the 
realizations $Q'_i$.

In the remainder of this paper, we consider scaled inclusivity because this 
method best captures the consistency and change in community organization.

\subsection{Generate weighted average maps of consistency}
Bias is inherent in the scaled inclusivity of a network that is based on a 
single referent partition $Q_R$. Every other partition is compared to this 
partition; the classification of the nodes in $Q_R$ has significant effects on 
the consistency values. To minimize this bias, the best partition for each 
realization ($Q'_i$ for $i=1,2,\cdots,n$) is selected as the referent partition 
$Q_R$ in turn. Thus, $n$ maps are computed. These maps are then averaged 
together using group similarity weights from the Jaccard similarity index. This 
serves to minimize the bias of any one network while still valuing the 
partitions that are most similar to the rest of the group.

Recall that matrix $J$ contains the Jaccard similarity index for every pair of 
partitions $Q'_i$ and $Q'_j$, with $J_{ij}$=0 where $i = j$. The 
column sum of $J$, $J_i=\sum_{j=1}^n J_{ij}$, is calculated, and the resulting 
vector is normalized to give the weights for the weighed average. Thus, the 
weight assigned to each scaled inclusivity map is proportional to the summed 
similarity of that partition to the rest of the group.

\subsection{Determine most representative partition by node}
A further analysis of scaled inclusivity can provide information about which 
realization is best when evaluating a certain portion of the network. To 
calculate this, a scaled inclusivity map is generated with each partition in 
turn as the referent partition, as above. From these, a map can be generated to 
show, for each node in the network, which realization has the highest scaled 
inclusivity value. This indicates that the node's classification in that 
realization is its most consistent classification - that the group community 
structure in this region is best captured by this partition. Groups of nodes 
that are best characterized by the same partition are of interest, especially 
if they correspond to a single community, because this indicates that this 
partition does a particularly good job of characterizing that community for 
the  entire group of networks. However, it should be noted that the actual 
scaled inclusivity values should be taken into account: if a given area has low 
scaled inclusivity values, then it does not matter what partition best 
represents this region because it is inconsistent across all realizations.

\subsection{Generate more informative maps of individual communities}
Scaled inclusivity, as described above, has a major limitation when considering 
a single community from a single partition. Such an analysis only includes 
nodes in the community of interest, even if every other partition had more 
nodes in the corresponding community. To solve this problem, another map can be 
made separately for each community in $Q_R$. 
The first step is to add the value $X_{iR}^{pq}$ to $A_p \cap R_q$, just as 
above. Secondly, 
the value $X_{iR}^{pq}$ is subtracted from every node in $A_p$ that is not in 
$R_q$ ($A_p \cap R_q^C$). Thus, if $A_p$ and $R_q$ have any overlap, nodes in 
the intersection will have a certain value added, and nodes only in $A_p$ will 
have a negative value with equal magnitude subtracted. These values can be 
summed for $p=1,2,\cdots,y$ in $Q'_i$ where $i=1,2,\cdots,n$ and $Q'_i \neq 
Q_R$ as long as $R_q$ is held constant. In this way, the negative values will 
not cancel out positive values - positive values occur only for nodes in $R_q$, 
and negative values occur only for nodes not in $R_q$. Much as high positive 
values indicate consistent classification in $R_q$ and a community in $Q'_i$ 
with significant overlap with $R_q$, very negative values indicate consistent 
classification outside of $R_q$ but in a community in $Q'_i$ with significant 
overlap with $R_q$. In other words, they are consistently classified in the 
same community as nodes in $R_q$ across the other partitions. This is a useful 
distinction to make in the case the referent partition contains a community of 
interest that lacks certain nodes or segments that are included in all other 
partitions. In this case, the values of those nodes would be very negative, 
indicating group consistency in spite of absence from the referent partition's 
community.


\section{Data}

\subsection{Simulated Data}

The first data set to be tested consists of 30 simulated networks using code 
from \cite{benchmark}. These networks were mapped onto a network of the 256 
most populous cities in the United States to more easily visualize the 
community structure of the network.

Lancichinetti, et al. \cite{benchmark} describe a method for creating 
unweighted, undirected networks to test various community structure algorithms. 
A key component of this algorithm is that degree and community size 
distributions are power laws, as in many real-world networks.

Parameters of the algorithm are number of nodes ($N$), mixing parameter 
($\mu$), average degree ($k_{ave}$), maximum degree ($k_{max}$), minimum 
community size ($s_{min}$), maximum community size ($s_{max}$), and the 
exponents of the power law degree and community size distributions ($\gamma$ 
and $\beta$, respectively). The mixing parameter $\mu$ is defined such that 
every node shares a fraction 1-$\mu$ links with other nodes in its community 
and a fraction $\mu$ links with nodes outside its community.

The algorithm computes degree and community size distributions based on the 
input parameters, and nodes are assigned a degree. Nodes are then assigned to 
communities, and the network is rewired to preserve degree and approximate 
$\mu$. 

The code to run this algorithm was made freely available online by the authors 
 (http://santo.fortunato.googlepages.com/benchmark. tgz). We used this code to 
generate 30 networks, each with $N$ = 256, $\mu$ = 0.35, $k_{ave}$ = 10, 
$k_{max}$ = 50, $s_{min}$ = 15, $s_{max}$ = 51, $\gamma$ = 2, and $\beta$ = 1. The network 
size was chosen 
to be fairly small for computational ease. Minimum community size was set to 
give a relatively consistent number of communities ($m$). The remaining 
parameters were adjusted slightly from the default values such that community 
analysis would show imperfect results similar to the community analysis of 
real-world networks. The sizes of the communities for each of the 30 networks 
can be seen in Table \ref{sim_communities}.

\begin{table}
\caption{Community sizes for all 30 simulated networks, listed in ascending 
order, shown to give the reader some idea about the overall community 
structure. 
\label{sim_communities}}
\begin{ruledtabular}
\begin{tabular}{ c | c c c c c c c c c c c c}
Network & \multicolumn{12}{c}{Number of Nodes in each Community} \\
\hline
1 & 16 & 16 & 17 & 29 & 29 & 32 & 33 & 39 & 45 \\
2 & 15 & 15 & 19 & 20 & 22 & 29 & 30 & 36 & 70 \\
3 & 15 & 18 & 21 & 27 & 30 & 34 & 35 & 38 & 38 \\
4 & 15 & 15 & 18 & 19 & 21 & 22 & 22 & 24 & 30 & 32 & 38 \\
5 & 15 & 18 & 22 & 23 & 27 & 29 & 34 & 40 & 48 \\
6 & 15 & 15 & 17 & 19 & 24 & 27 & 29 & 31 & 35 & 44 \\
7 & 15 & 18 & 20 & 22 & 23 & 27 & 27 & 28 & 34 & 42 \\
8 & 22 & 22 & 26 & 28 & 29 & 31 & 31 & 31 & 36 \\
9 & 15 & 15 & 16 & 20 & 24 & 30 & 32 & 33 & 34 & 37 \\
10 & 20 & 25 & 26 & 28 & 28 & 29 & 32 & 33 & 35 \\
11 & 16 & 16 & 18 & 19 & 20 & 20 & 21 & 21 & 32 & 36 & 37 \\
12 & 17 & 23 & 28 & 31 & 36 & 38 & 39 & 44 \\
13 & 16 & 22 & 23 & 30 & 31 & 39 & 40 & 55 \\
14 & 17 & 17 & 19 & 24 & 26 & 27 & 38 & 40 & 48 \\
15 & 16 & 21 & 26 & 26 & 31 & 31 & 33 & 36 & 36 \\
16 & 15 & 19 & 20 & 20 & 22 & 24 & 28 & 33 & 35 & 40 \\
17 & 17 & 22 & 24 & 28 & 35 & 35 & 38 & 57 \\
18 & 19 & 22 & 24 & 24 & 27 & 30 & 35 & 36 & 39 \\
19 & 15 & 21 & 26 & 31 & 31 & 32 & 32 & 32 & 36 \\
20 & 22 & 30 & 32 & 32 & 32 & 33 & 37 & 38 \\
21 & 17 & 18 & 19 & 20 & 24 & 25 & 27 & 32 & 35 & 39 \\
22 & 16 & 16 & 19 & 20 & 20 & 21 & 23 & 33 & 38 & 50 \\
23 & 16 & 17 & 19 & 26 & 28 & 33 & 38 & 39 & 40 \\
24 & 17 & 17 & 17 & 18 & 18 & 18 & 18 & 23 & 25 & 26 & 27 & 32 \\
25 & 15 & 16 & 17 & 17 & 18 & 22 & 25 & 33 & 34 & 59 \\
26 & 15 & 15 & 16 & 17 & 18 & 24 & 26 & 29 & 30 & 32 & 34 \\
27 & 20 & 23 & 31 & 32 & 33 & 35 & 36 & 46 \\
28 & 15 & 19 & 20 & 25 & 25 & 26 & 27 & 29 & 29 & 41 \\
29 & 18 & 22 & 24 & 25 & 27 & 27 & 27 & 40 & 46 \\
30 & 16 & 19 & 19 & 26 & 29 & 33 & 34 & 40 & 40 \\ 
\end{tabular}
\end{ruledtabular}
\end{table}

\begin{figure}[h]
\begin{center}
\includegraphics[width=80mm,height=40mm]{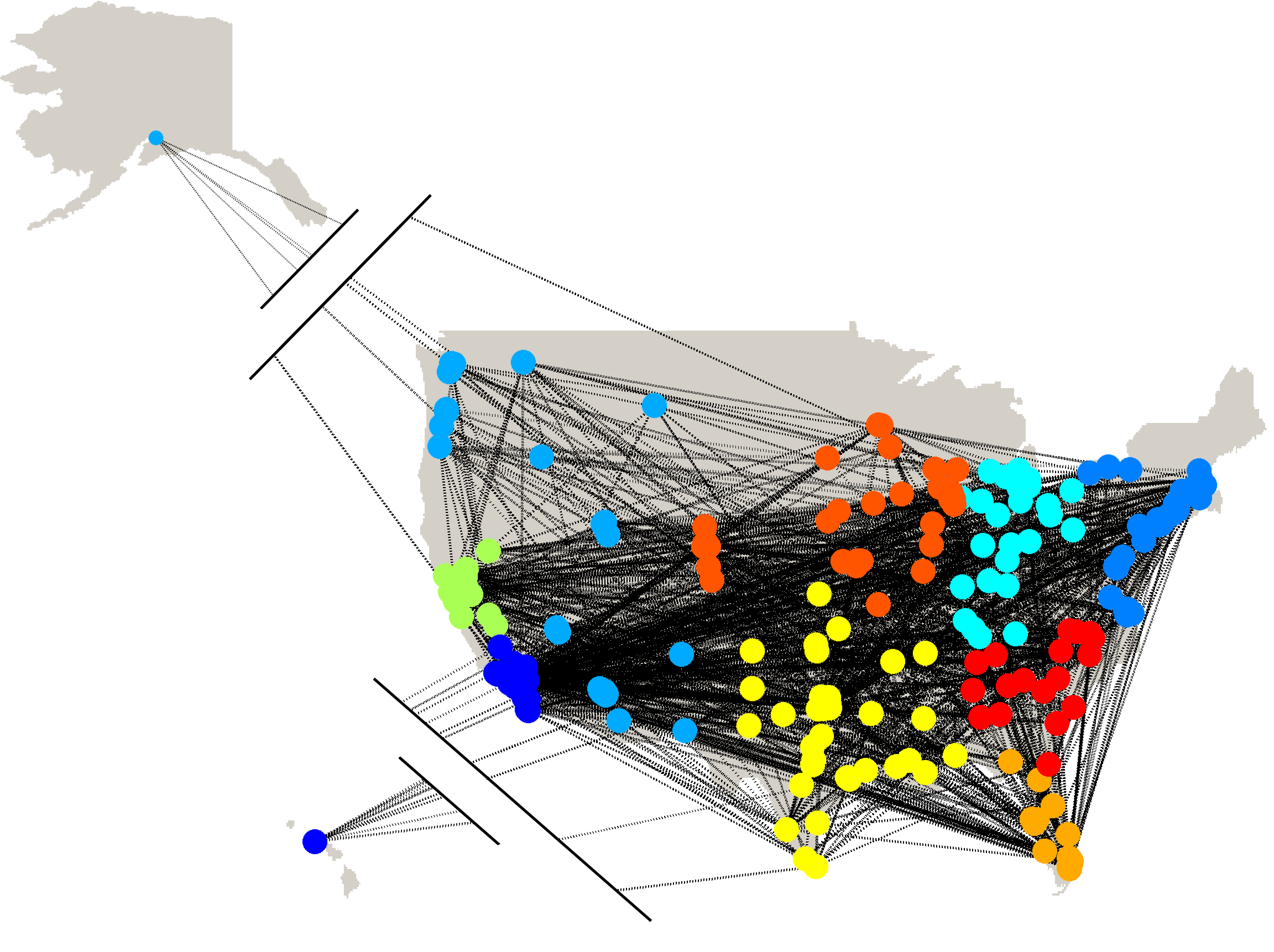} \\
\end{center}
\caption{(Color online) Network 30, with the nodes color-coded by simulated 
community. 
\label{network}}
\end{figure}

The nodes assigned to a given community in any individual network were random. 
In order to visualize these networks and to give the networks some consistency 
in their community structure, the nodes were assigned to the 256 most populous 
cities in the United States according to 2009 census data \cite{census}. 
Consider the matrix $C$, whose rows $c_i$ ($i=1,2,\cdots,256$) contain the 
longitude and latitude of a given city $i$ ($i$ = 1,2,$\cdots,256$). Recall 
that $A_j$ is the set of nodes in community $j$ ($j$ = 1,2,$\cdots$,$m$) and 
$s_j$ = $|A_j|$. Let $C_j$ be the set of coordinate pairs $c_i$ ($i \in$ 
\{1,2,$\cdots$,256\}) such that $C_j$ also has size $s_j$. For each network, 
each city was manually assigned to a community based on community size $s_j$ 
and geographical proximity. Each node $n_l \in A_j$ is assigned to an arbitrary 
node $c_i \in C_j$, and the cities inherit the links from the associated nodes 
to keep the network structure intact.

\begin{figure}[h]
\begin{center}
\subfigure[Simulated Communities for Network 30]{\label{fig:edge-a}
\includegraphics[width=80mm,height=40mm]
{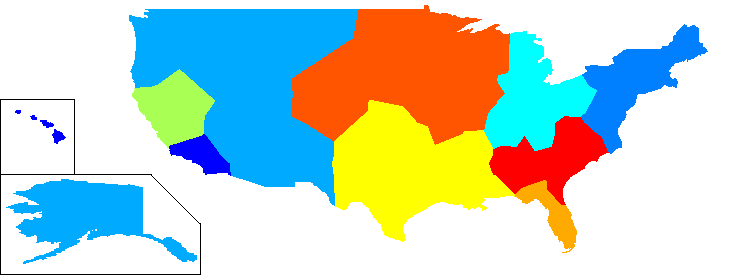}}
\subfigure[Simulated Communities for Network 3]
{\label{fig:edge-b}\includegraphics[width=80mm,height=40mm]
{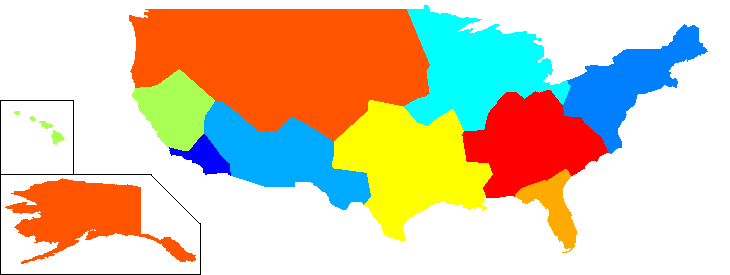}} \\
\end{center}
\caption{(Color online) Simulated communities for network 30 (a) and simulated 
communities from network 3 for comparison (b). Here, the colors indicate 
membership in a community. \label{net30}}
\end{figure}

An example of one of the networks (network 30) is shown in Fig. \ref{network} 
with each node color-coded according to community membership. We also generated 
an alternative space-filling visualization. The United 
States Census Bureau releases the latitude and longitude of the borders of the 
country \cite{census}. A map was generated using these coordinates, and it was 
converted into a 1200x600 image. Let $p_{xy}$ be a pixel in this grid ($x \in$ 
\{1,2,$\cdots$, 1200\}, $y \in$ \{1,2,$\cdots$,600\}). For every $p_{xy}$ 
within U.S. borders, the Euclidean distance to each of the 256 most populous 
cities was computed, and the pixel was assigned to the nearest city. 
Fig. \ref{net30} shows the simulated communities for network 30 (a) and 
for network 3 (b). The communities are shown by color-coding the pixels 
assigned to each node. These networks are shown as examples of typical 
community  structure. The two are fairly similar, but no community is identical 
between them. It should be noted that due to differences in spacing, 
some nodes have very few assigned pixels, while others have many. In other 
words, the area of a community is not necessarily indicative of the number of 
nodes it contains.


\subsection{Real Data}

The second set of test data comes from NCAA College Football from the years 
1995 to 2009. In this network, each team that was active in the Football Bowl 
Subdivision (formerly Division IA, hereafter called FBS) constitutes a node, 
and every game played between two FBS teams is a link. In each season, teams 
play eleven or twelve regular season games and one or two postseason games. 
Teams are organized into conferences, which usually consist of between eight 
and twelve teams. Teams generally play other teams in their own 
conference more often than teams in other conferences. 
The main exceptions are independent schools, which are not members of any 
conference. In addition, schools may play one game each season against teams 
that are not in the FBS. These games were not counted in the network. In rare 
cases, teams may have played two games against the same opponent in one season. 
The second game was not counted; there were no weighted links in these 
networks. Some 
teams joined the FBS during the time period in question. Any games played by 
these teams against FBS opponents prior to joining the FBS were not counted in 
the network. For consistency in the analysis, these nodes were included in the 
networks for these seasons, but each team belonged to an isolated community 
until the year it joined the FBS.

This data set lends itself well to network analysis both because of its size 
(N=120) and because the communities in the network are well-defined in reality. 
Each conference represents a community, and each independent team is counted as 
a separate community as well.

\begin{figure}[h]
\begin{center}
\subfigure[True Conference Alignment in 2009]{\label{fig:edge-a}
\includegraphics[width=65mm,height=35mm]
{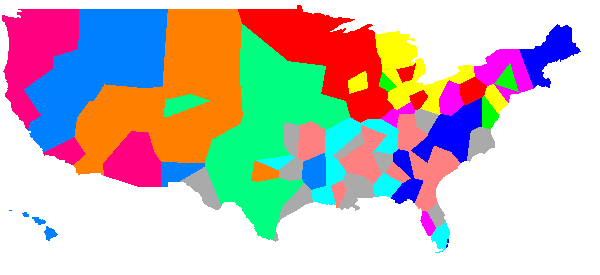}} \includegraphics[width=15mm,height=35mm]
{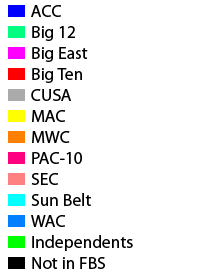}
\subfigure[True Conference Alignment in 2003]{\label{fig:edge-b}
\includegraphics[width=65mm,height=35mm]
{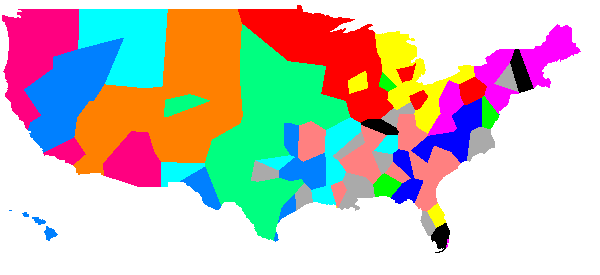}} \includegraphics[width=15mm,height=35mm]
{Legend_fig5.png}\\
\end{center}
\caption{(Color online) Maps showing the true conference alignment for FBS in 
2009 (a) and the 
true conference alignment for FBS in 2003 (b). The conferences are similar in 
these two years, but several teams move between conferences, and several more 
join the FBS.
\label{2009}}
\end{figure}

The conferences that existed in 2009 were the Atlantic Coast Conference (ACC), 
the Big 12, the Big East, the Big Ten, Conference USA (CUSA), the Mid-American 
Conference (MAC), the Mountain West Conference (MWC), the Pac-10, the 
Southeastern Conference (SEC), the Sun Belt Conference, and the Western 
Athletic Conference (WAC). In 2009, there were three independent teams: 
Notre Dame, Army, and Navy. Several conferences that existed in 1995 (the 
first year of our analysis) are no longer in existence: the Big 8 (now the Big 
12), the Big West, and the Southwest Conference (SWC). Finally, there were 120 
FBS teams in 2009, whereas there were 107 in 1995. The conference alignment for 
2009 and 2003 are shown in Fig. \ref{2009} as examples of the full landscape. 
All the same conferences existed in these two seasons, but a number of teams 
changed conferences, and four teams joined the FBS between 2003 and 2009.

The total number of teams in each conference for each season can be seen in 
Table \ref{real_communities}. Table \ref{real_communities} reveals some basic 
information about the consistency of community structure between 1995 and 2009.
For example, several conferences came into existence (CUSA, the MWC, and the 
Sun Belt), and several ceased to exist (the Big 8, the Big West, and the SWC). 
In addition, some conferences changed in size (the ACC, the Big East, the MAC, 
and the WAC), while others remained constant (the Big Ten, the Pac 10, and the 
SEC). Finally, the number of independents decreased fairly steadily over the 
time period in question.

\begin{table*}
\caption{Conference sizes for all 15 years of the the college football 
network,  shown to give the reader some idea about the overall community 
structure. 
\label{real_communities}}
\begin{ruledtabular}
\begin{tabular}{ | c | c | c | c | c | c | c | c | c | c | c | c | c | c | 
c | c |}
Year & ACC & Big 8 & Big 12 & Big East & Big Ten & Big West & CUSA & MAC & MWC 
& Pac 10 & SEC & SWC & Sun Belt & WAC & Ind.\\
\hline
1995 & 9 & 8 & 0 & 8 & 11 & 9 & 0 & 10 & 0 & 10 & 12 & 8 & 0 & 10 & 12\\
1996 & 9 & 0 & 12 & 8 & 11 & 6 & 6 & 10 & 0 & 10 & 12 & 0 & 0 & 16 & 11\\
1997 & 9 & 0 & 12 & 8 & 11 & 6 & 7 & 12 & 0 & 10 & 12 & 0 & 0 & 16 & 9\\
1998 & 9 & 0 & 12 & 8 & 11 & 6 & 8 & 12 & 0 & 10 & 12 & 0 & 0 & 16 & 8\\
1999 & 9 & 0 & 12 & 8 & 11 & 7 & 9 & 13 & 8 & 10 & 12 & 0 & 0 & 8 & 7\\
2000 & 9 & 0 & 12 & 8 & 11 & 6 & 9 & 13 & 8 & 10 & 12 & 0 & 0 & 9 & 9\\
2001 & 9 & 0 & 12 & 8 & 11 & 0 & 10 & 13 & 8 & 10 & 12 & 0 & 7 & 10 & 7\\
2002 & 9 & 0 & 12 & 8 & 11 & 0 & 10 & 14 & 8 & 10 & 12 & 0 & 7 & 10 & 6\\
2003 & 9 & 0 & 12 & 8 & 11 & 0 & 11 & 14 & 8 & 10 & 12 & 0 & 8 & 10 & 4\\
2004 & 11 & 0 & 12 & 7 & 11 & 0 & 11 & 14 & 8 & 10 & 12 & 0 & 11 & 10 & 2\\
2005 & 12 & 0 & 12 & 8 & 11 & 0 & 12 & 12 & 9 & 10 & 12 & 0 & 8 & 9 & 4\\
2006 & 12 & 0 & 12 & 8 & 11 & 0 & 12 & 12 & 9 & 10 & 12 & 0 & 8 & 9 & 4\\
2007 & 12 & 0 & 12 & 8 & 11 & 0 & 12 & 13 & 9 & 10 & 12 & 0 & 8 & 9 & 4\\
2008 & 12 & 0 & 12 & 8 & 11 & 0 & 12 & 13 & 9 & 10 & 12 & 0 & 9 & 9 & 3\\
2009 & 12 & 0 & 12 & 8 & 11 & 0 & 12 & 13 & 9 & 10 & 12 & 0 & 9 & 9 & 3\\
\hline
\end{tabular}
\end{ruledtabular}
\end{table*}

The mapping procedure for the simulated data was replicated here with 
$c_i$ containing the latitude and longitude of team $i$ 
($i$ = 1,2,$\cdots$,120).
Thus, the pixels were assigned to teams instead of cities. As before, the 
number of pixels assigned to a single node varied greatly; community size in 
pixels does not necessarily indicate community size in number of teams.

\section{Results}

\subsection{Simulated Data}
\begin{figure}[h]
\begin{center}
\includegraphics[width=80mm,height=40mm]{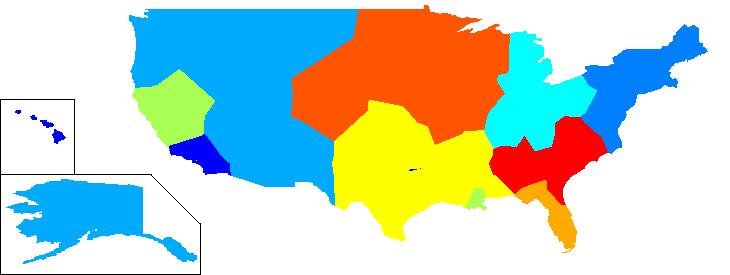} \\
\end{center}
\caption{(Color online) Network 30, with the nodes color-coded by community 
according to the best run of QCut. Note the misclassified nodes in Texas and 
Louisiana.
\label{net_30_QCut}}
\end{figure}

We conducted the analysis in Section II on the 30 simulated networks to 
evaluate the effectiveness of the algorithm. For this analysis, we ran QCut 10 
times for each realization ($g=10$). The community sizes of the best results of 
QCut can be seen in Table \ref{sim_communities_QCut}, included for comparison 
with the simulated community sizes as seen in Table \ref{sim_communities}. 
Here, we will look more closely at the scaled inclusivity of a representative 
network (network 30). The 
simulated communities can be seen in Fig. \ref{net30}(a), and the best QCut 
results for network 30 can be seen in Fig. \ref{net_30_QCut}. Note the 
misclassified nodes in the community containing Texas in the QCut results 
(which are to be expected from any community structure algorithm).

\begin{table}
\caption{Community sizes and summed Jaccard index values for the QCut results 
of all 30 simulated networks, listed in ascending order, to show the changes in 
community structure due to QCut. 
\label{sim_communities_QCut}}
\begin{ruledtabular}
\begin{tabular}{ c | c | c c c c c c c c c c c c}
Network & $J'_i$ & \multicolumn{9}{c}{Community Sizes} \\
\hline
1 & 14.1 & 16 & 17 & 22 & 28 & 31 & 32 & 33 & 34 & 43 \\
2 & 13.4 & 15 & 17 & 18 & 20 & 23 & 29 & 30 & 36 & 68 \\ 
3 & 15.0 & 16 & 18 & 21 & 28 & 32 & 32 & 36 & 36 & 37 \\
4 & 14.2 & 15 & 15 & 17 & 20 & 20 & 22 & 22 & 24 & 30 & 32 & 39  \\
5 & 14.3 & 15 & 18 & 22 & 23 & 29 & 29 & 33 & 42 & 45  \\
6 & 14.8 & 15 & 15 & 17 & 19 & 24 & 27 & 29 & 31 & 35 & 44  \\
7 & 15.3 & 16 & 18 & 21 & 22 & 22 & 27 & 27 & 28 & 34 & 41  \\
8 & 14.7 & 22 & 22 & 26 & 28 & 29 & 31 & 31 & 31 & 36    \\
9 & 15.4 & 15 & 16 & 17 & 20 & 22 & 30 & 32 & 33 & 34 & 37  \\
10 & 14.8 & 20 & 27 & 27 & 28 & 28 & 28 & 31 & 32 & 35  \\
11 & 14.7 & 16 & 16 & 18 & 19 & 20 & 20 & 21 & 21 & 32 & 36 & 37     \\
12 & 13.7 & 17 & 24 & 28 & 31 & 36 & 38 & 40 & 42  \\
13 & 14.1 & 16 & 23 & 23 & 30 & 31 & 39 & 41 & 53   \\
14 & 15.4 & 17 & 18 & 19 & 25 & 25 & 27 & 38 & 39 & 48 \\ 
15 & 13.3 & 18 & 21 & 27 & 27 & 31 & 32 & 32 & 32 & 36    \\
16 & 14.5 & 15 & 19 & 20 & 20 & 22 & 24 & 28 & 33 & 35 & 40  \\
17 & 14.3 & 17 & 24 & 28 & 28 & 35 & 35 & 38 & 51  \\
18 & 14.0 & 19 & 22 & 24 & 24 & 27 & 32 & 35 & 36 & 37     \\
19 & 15.0 & 15 & 21 & 26 & 30 & 31 & 32 & 32 & 33 & 36    \\
20 & 13.5 & 22 & 30 & 32 & 32 & 33 & 34 & 36 & 37  \\
21 & 15.4 & 17 & 18 & 19 & 20 & 24 & 25 & 27 & 32 & 35 & 39 \\
22 & 13.5 & 16 & 16 & 19 & 20 & 21 & 21 & 23 & 33 & 38 & 49  \\
23 & 15.2 & 17 & 17 & 21 & 27 & 28 & 33 & 36 & 37 & 40  \\
24 & 13.0 & 17 & 17 & 17 & 18 & 18 & 18 & 19 & 23 & 25 & 26 & 26 & 32 \\
25 & 12.6 & 17 & 17 & 17 & 18 & 19 & 22 & 25 & 33 & 34 & 54  \\
26 & 14.1 & 15 & 15 & 16 & 17 & 18 & 24 & 26 & 29 & 30 & 32 & 34 \\
27 & 14.5 & 21 & 24 & 30 & 32 & 33 & 35 & 36 & 45  \\
28 & 15.2 & 15 & 19 & 20 & 25 & 25 & 26 & 27 & 29 & 30 & 40 \\
29 & 14.5 & 18 & 22 & 24 & 25 & 27 & 27 & 27 & 41 & 45 \\
30 & 15.5 & 16 & 19 & 20 & 26 & 29 & 33 & 34 & 39 & 40 \\\end{tabular}
\end{ruledtabular}
\end{table}

\begin{figure}[h]
\begin{center}
\subfigure[Scaled Inclusivity Across Simulated Communities]{\label{fig:edge-a}
\includegraphics[width=75mm,height=40mm]
{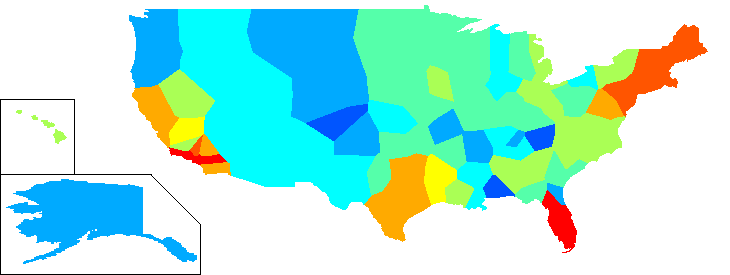}} \includegraphics[width=5mm,height=40mm]
{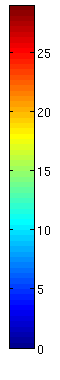}
\subfigure[Scaled Inclusivity Across QCut Results]{\label{fig:edge-b}
\includegraphics[width=75mm,height=40mm]
{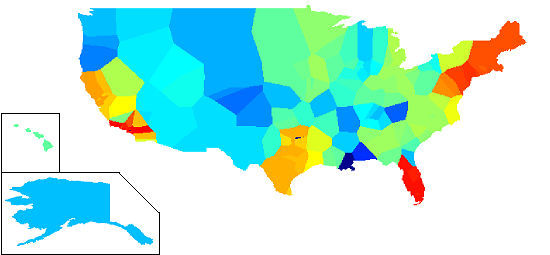}} \includegraphics[width=5mm,height=40mm]
{colorbar29_fig7.png} \\
\end{center}
\caption{(Color online) Scaled inclusivity maps generated using simulated 
communities (a) and the QCut results (b). For both maps, network 30 was used as 
the referent partition (maps in Fig. \ref{net30}). Note the effects of the 
misclassifications visible in Fig. \ref{net_30_QCut} on (b) here. \label{con30}}
\end{figure}

\begin{figure}[h]
\begin{center}
\subfigure[SI for Simulated Texas Community]{\label{fig:edge-a}
\includegraphics[width=37mm,height=37mm]{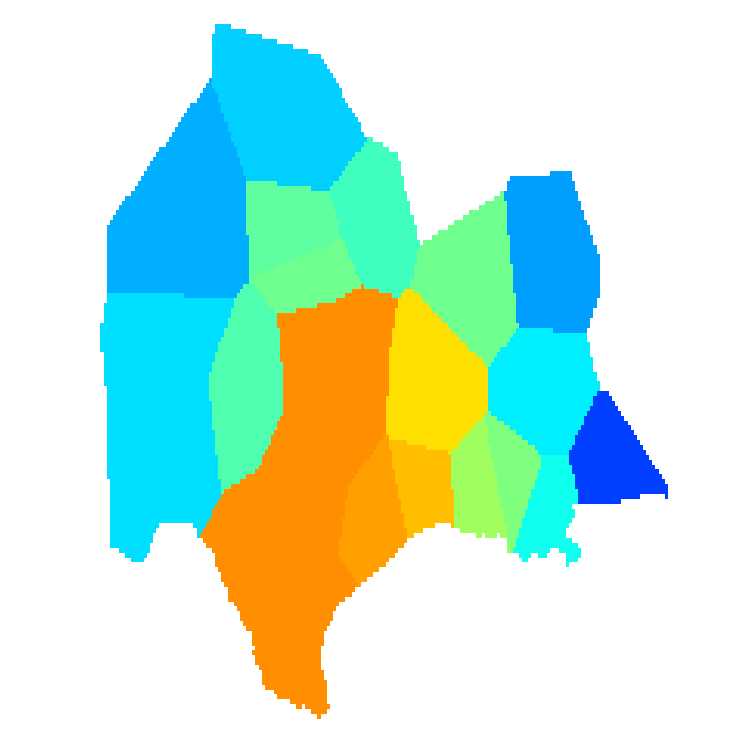}} 
\subfigure[SI for QCut Results on Simulated Texas Community]{\label{fig:edge-b}
\includegraphics[width=37mm,height=37mm]{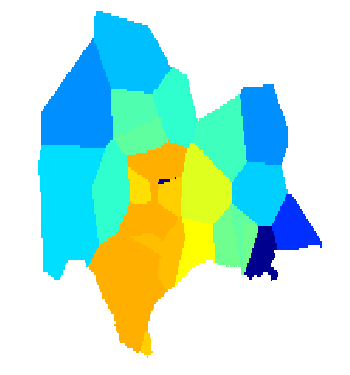}} 
\includegraphics[width=5mm,height=37mm]{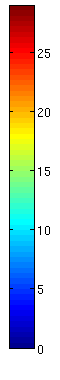}\\
\end{center}
\caption{(Color online) Scaled inclusivity for the simulated community 
including Texas for simulated communities and the results of QCut. As before, 
both maps were generated with network 30 as the referent partition. Note the 
similarity of results in some regions and the vast differences for the two 
misclassified nodes from Fig. \ref{net_30_QCut}. 
\label{Texas}}
\end{figure}

\begin{figure}[h]
\begin{center}$
\begin{array}{cc}
\includegraphics[width=70mm,height=70mm]{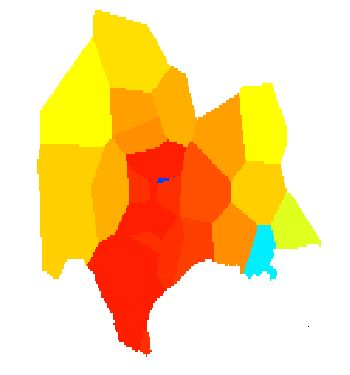} &
\includegraphics[width=10mm,height=70mm]{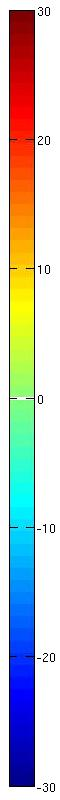}
\end{array}$
\end{center}
\caption{(Color online) Scaled inclusivity including negative values for the 
simulated community including Texas using results from QCut, generated using 
network 30 as the referent partition. Note that the 
central node that was misclassified is highly negative, indicating the 
likelihood that it is consistently in the same module as the nodes surrounding 
it for most other realizations in the group. The node in Louisiana that also 
has negative values has a much smaller absolute value, suggesting that it was 
not consistent in its classification across the other realizations.
\label{SI_neg}}
\end{figure}

\begin{figure}[h]
\begin{center}$
\begin{array}{cc}
\includegraphics[width=75mm,height=40mm]{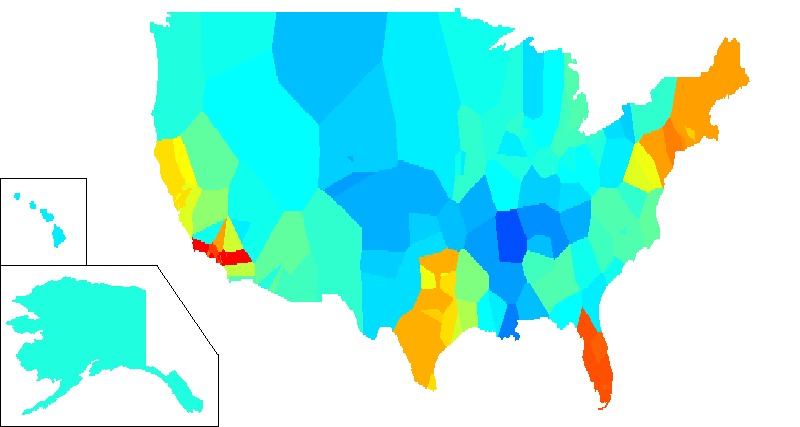} &
\includegraphics[width=5mm,height=40mm]{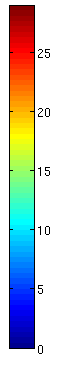}
\end{array}$
\end{center}
\caption{(Color online) Weighted average of scaled inclusivity maps across all 
communities using QCut results. Note that the misclassified nodes do not 
significantly contribute to the 
final values for those nodes from any network when all possible community 
structures are considered. This indicates that the nodes are in consistent 
neighborhoods in all networks except the referent network (network 30) 
used  in the analysis. \label{SIAW}}
\end{figure}

The scaled inclusivity map using network 30 as the reference partition $Q_R$ 
was generated for both the simulated communities and for the results of QCut 
(Fig. \ref{con30}). Notice the similarities and 
differences between the two maps, especially for the two nodes that were 
misclassified in network 30 (as seen in Fig. \ref{net_30_QCut}). In the map 
made from the partitions generated by QCut, those nodes 
were not consistently classified in the communities they occupy in network 30, 
so their scaled inclusivity values were quite low. This would seem to indicate 
that these nodes were not at all consistently classified across networks. 
However, in the simulated communities, these nodes were fairly consistent in 
their classification. The community including Texas was isolated for both the 
simulated communities and the results of QCut (Fig. \ref{Texas}) to make the 
differences more easily visible. Note that the nodes in central and south 
Texas have the highest scaled inclusivity values, and the nodes around that 
region have lower values. The appendix contains images of all simulated 
communities that overlap with the Texas community for each realization to allow 
the reader to visually evaluate the consistency of the nodes in that community. 
The nodes in this region were classified together in the same community in all 
other networks, but due to imperfect overlap between their communities and the 
community shown here, their scaled inclusivity values are less than the optimal 
value of 29. The other nodes shown here are sometimes included in that 
community and sometimes not, resulting in their lower scaled inclusivity values.

In addition, the scaled inclusivity map containing both positive and negative 
values is shown in Fig. \ref{SI_neg}. This map appears very similar to Fig. 
\ref{Texas}(b), but the negative values reveal more information. The node in 
central Texas has a very negative value - its absolute value is roughly equal 
to the nodes around it with positive values. The node in Louisiana, 
however, is not so negative. This indicates that the node in central Texas is 
consistently classified with the other nodes around it in the other 
realizations of the network; the node in Louisiana, though, is inconsistent in 
its classification across realizations. Thus, for the group, we can look at the 
node in Texas as belonging to the community containing the rest of central and 
south Texas across most subjects. The negative values reveal the distinction 
between the two misclassified nodes in this community.

In a real data set, the true community structure is typically unknown; the only 
data available are the results of a community detection algorithm, making 
misclassified nodes such as 
those in Texas and Louisiana much more difficult to detect. Nodes that were 
misclassified or merely in a different community only in the referent partition 
appear to have very low scaled inclusivity values, as seen in Fig. 
\ref{Texas}(b). However, in part (a), it can be seen that those nodes were 
more consistently classified in the simulated communites. In order to more 
effectively evaluate the consistency of each node, irrespective of the specific 
community assignment, we generated the weighted average map, which can be 
seen in Fig. \ref{SIAW}. The Jaccard index value for each network, which was 
used to generate the weights for the weighted average, 
is included in Table \ref{sim_communities_QCut}. Note that 
the nodes that were misclassified in network 30 have values that better match 
their neighbors and better reflect their consistency across all 
networks. This visualization may give better information about the consistency 
of individual nodes across multiple realizations when the true community 
structure is unknown. There is one significant 
caveat to this approach. The weighted average uses a partition from every 
realization as a referent. As a result, there is no single underlying community 
structure, and its consistency cannot be evaluated. However, the consistency 
of each individual node is better characterized.

\begin{figure}[h]
\begin{center}$
\begin{array}{cc}
\includegraphics[width=75mm,height=40mm]{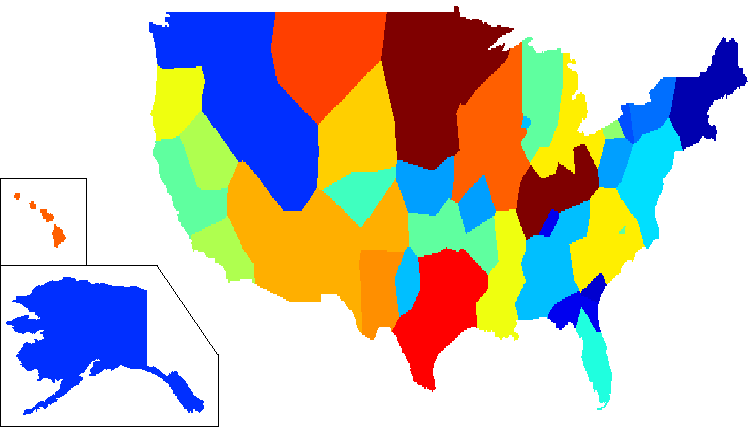} &
\includegraphics[width=5mm,height=40mm]{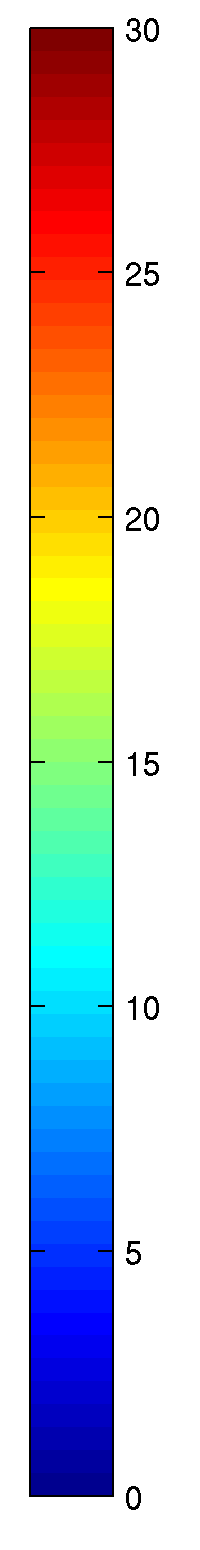}
\end{array}$
\end{center}
\caption{(Color online) Map showing, for each node, the index of the 
realization that, when used as the reference partition, yielded the highest 
scaled inclusivity value. In other words, this indicates which subject's scaled 
inclusivity map is most representative for each node across the entire group.
\label{max_SI}}
\end{figure}

\begin{figure}[h]
\begin{center}$
\begin{array}{cc}
\includegraphics[width=55mm,height=70mm]{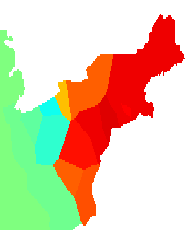} &
\includegraphics[width=10mm,height=70mm]{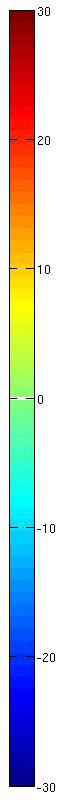}
\end{array}$
\end{center}
\caption{(Color online) Scaled inclusivity including negative values for the 
simulated community including the northeastern United States, using partition 
1 as the referent partition $Q_R$. Most of the green (light gray) values in 
this 
image are very close to 0; some nodes farther west had very small values, but 
they were not pictured for space reasons.
\label{NE_neg}}
\end{figure}

Fig. \ref{max_SI} shows another map of the USA. In this case, every partition 
was used as the referent partition $Q_R$, and a map was made showing which 
partition yielded the highest scaled inclusivity value for each node. This map 
reveals which referent partition best characterizes each node or region.This 
map is useful when interpreting the negative scaled inclusivity maps together 
with the weighed average map. Regions of consistent community structure 
can be identified on the weighted average map. To then evaluate the 
organization of a community in any particular region, one would consult a 
negative scaled inclusivity map. However, as negative scaled inclusivity maps, 
by necessity can only be generated for one community of a particular referent 
partition, the "best" referent needs to be identified. Maps showing the 
referent that had the highest scaled inclusivity value (Fig. \ref{max_SI}) are 
then consulted to pick the most appropriate partition to examine. For 
example, the northeastern United States has relatively high values, and the 
region is best characterized by partition 1. The negative scaled 
inclusivity map for that region can be seen in Fig. \ref{NE_neg}.

\subsection{Real Data}

\begin{figure}[h]
\begin{center}
\includegraphics[width=80mm,height=40mm]{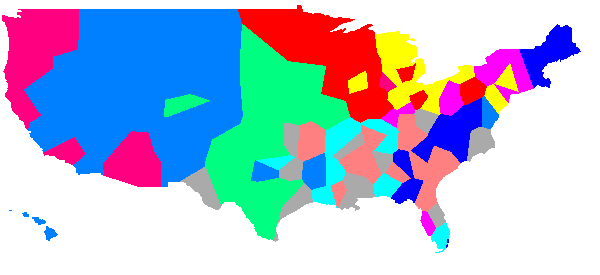} \\
\end{center}
\caption{(Color online) The best partition from the QCut runs for 2009.
\label{FBS_rep_QCut}}
\end{figure}

We again conducted the analysis as described in Section II (with $g=10$) to get 
the best partition from QCut ($Q'_i$) for every year under consideration 
for the college football networks. The community structure of the best run of 
QCut for 2009 is visible in Fig. \ref{FBS_rep_QCut}. It 
should be noted that this partition is not the same as the true conference 
alignment. The football conferences form well-defined communities in general, 
but a network based solely on games played may result in partitions that do not 
directly match the conferences. For example, independent schools are put into 
a larger community, in part because they tend to play a large number of schools 
from one conference. It some cases, teams from two geographically close 
conferences played a large number of interconference games. An optimal 
partition of this network lumps two such conferences together into one 
community. This occurred during multiple seasons with the Sun Belt and the SEC 
as well as with the MWC and the WAC.

\begin{figure}[h]
\begin{center}
\subfigure[Scaled Inclusivity for True Conference Alignment]{\label{fig:edge-a}
\includegraphics[width=78mm,height=40mm]{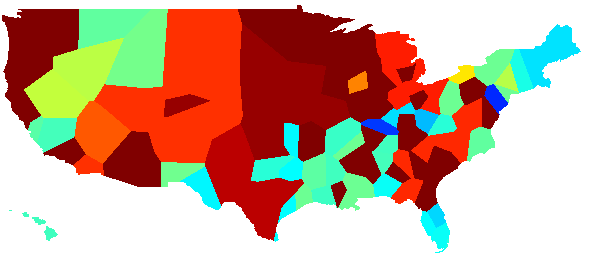}}
\includegraphics[width=5mm,height=40mm]{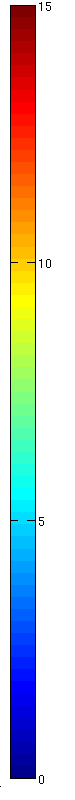}
\subfigure[Scaled Inclusivity for QCut Conference Alignment]{\label{fig:edge-b}
\includegraphics[width=78mm,height=40mm]{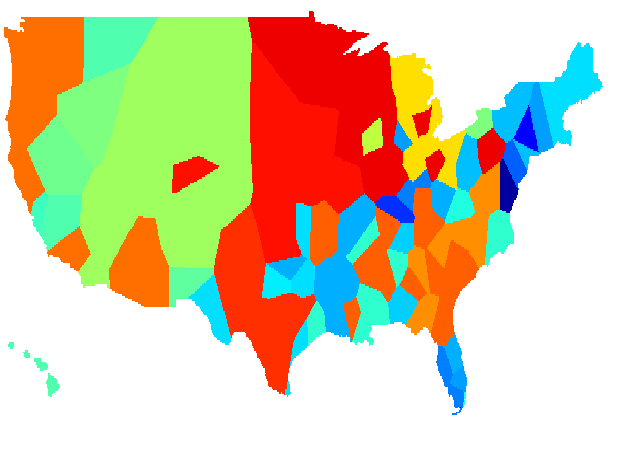}} 
\includegraphics[width=5mm,height=40mm]{colorbar15_fig14.png}\\
\end{center}
\caption{(Color online) Scaled inclusivity map for FBS conferences, with 2009 
as the referent network for comparison. Shown for true conference alignment in 
(a) and the QCut approximations in (b). \label{QSI2009}}
\end{figure}

The overall scaled inclusivity map was generated for the true conference 
alignment and the results of QCut, and the results are visible in Fig. 
\ref{QSI2009}. The consistency is significantly higher in the network showing 
the true conference alignment. This is due to differences between the true 
conference membership and the community structure identified using QCut.

\begin{figure}[h]
\begin{center}$
\begin{array}{cc}
\includegraphics[width=75mm,height=40mm]{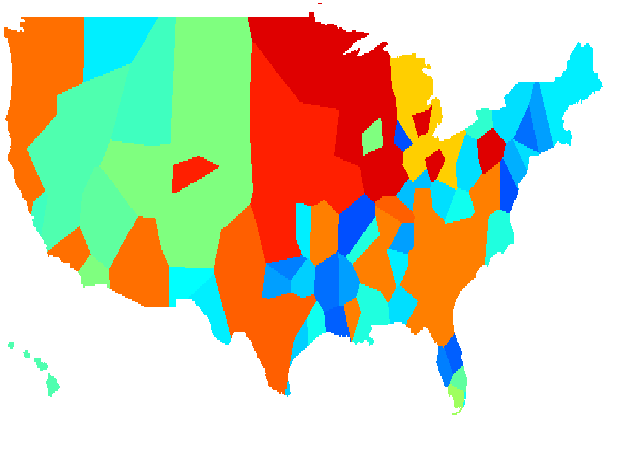} &
\includegraphics[width=5mm,height=40mm]{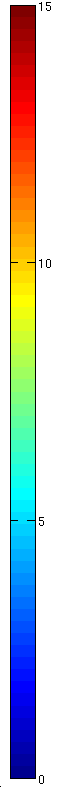}
\end{array}$
\end{center}
\caption{(Color online) Weighted average map for the scaled inclusivity of the 
FBS conferences, shown for the results of QCut.
\label{FBS_weighted_all}}
\end{figure}

\begin{figure}[h]
\begin{center}$
\begin{array}{cc}
\includegraphics[width=75mm,height=40mm]{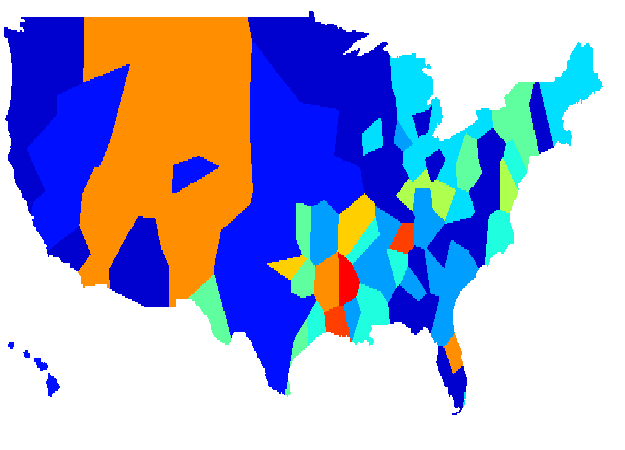} &
\includegraphics[width=5mm,height=40mm]{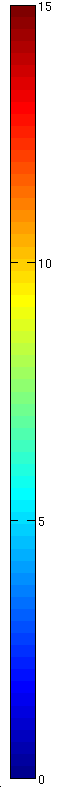}
\end{array}$
\end{center}
\caption{(Color online) Year with maximum scaled inclusivity for each node. 
Network 1 here represents 1995, and network 15 represents 2009.
\label{FBS_max_SI}}
\end{figure}

For time series data, there is likely a single network (probably the first or 
last realization, although not necessarily) that is of greatest interest. In 
networks with linear order, the scaled inclusivity for the most recent 
realization shows how consistently the community structure over time aligns 
with the current or most recent version of the network. However, the overall 
characterization of the group of networks may be of interest as well. The 
scaled inclusivity of the FBS networks was shown for 2009 because it is the 
most recent network; we have also included the weighted average map for FBS 
(Fig. \ref{FBS_weighted_all}) and the maximum scaled inclusivity map (Fig. 
\ref{FBS_max_SI}). Interestingly, the weighted average map is very similar to 
the map for 2009. Of even more interest is that the maximum scaled inclusivity 
values appeared mostly in the first few networks. It should be noted here that 
because of how this was calculated, in the event of equal maximum scaled 
inclusivity values, the earliest network in the time series will be chosen. 
This may be the case here for conferences that remained unchanged throughout 
the time period studied, such as the PAC 10 and the Big Ten.


\section{Discussion}

This paper has introduced scaled inclusivity, which is a new method for 
evaluating the consistency of the partitioning of related networks. 
There are several methods for determining the community structure of a network; 
the best way to partition a network is an area of ongoing research. The 
comparison of community analyses has recently become of interest as well. 
Emphasis has been placed on the change in community structure over time as 
various datasets evolve. 

Several methods have been proposed for the comparison of community structure in 
different networks. Palla, et. al. \cite{Palla} studied the relationships 
between size, age, stationarity, and lifetime of communities. They investigate 
the characteristics of communities with short and long lifespans in changing 
networks. The stationarity 
is defined as the average consistency from each time step to the next. They 
propose a method for determining the best match among communities in 
consecutive time steps; we consider every pair of communities with any overlap 
when calculating 
scaled inclusivity. Asur et. al. \cite{Asur} define community events (continue, 
merge, split, form, and dissolve) and node events (appear, disappear, join, and 
leave). They captured the occurrence of these events over time and used that 
information to characterize the dynamic community structure of the network. 
Chakrabarti, et. al. \cite{Chakrabarti} define a method for evolutionary 
clustering, which seeks to balance the quality of fit for any specific network 
with similarity to previous and subsequent partitions. Fenn, et. al. 
\cite{Fenn} track community changes by using centrality metrics 
to analyze the roles of individual vertices in economic networks. As in scaled 
inclusivity, two communities at different times are compared if they have any 
common nodes; the method does not attempt to select the best matching 
communities. These methods are designed specifically to characterize change 
over time but are not well suited to cross-sectional analysis.

Hopcroft, et. al. \cite{Hopcroft} take a different approach. They use a 
community detection algorithm that finds the "natural communities" of a 
network. This analysis leaves some nodes out as not belonging to any community. 
Then, to compare across time, they use a metric they call the best 
match, which is defined as follows for two communities $C$ and $C'$: 
\begin{equation} match(C,C') = min \left( \frac{|C \cap C'|}{|C|},
\frac{|C \cap C'|}{|C'|} \right). \end{equation} In that paper, they 
investigated citation networks from the NEC CiteSeer database at two different 
time points. Because citation networks never lose edges and because only two 
realizations were investigated, they manually compared the best match 
communities from the two datasets, noting the growth and emergence of 
communities over time and the corresponding changes in the fields they 
characterize. This method could be applied on longitudinal or cross-sectional 
datasets, provided that the number of networks is limited.

To our knowledge, this is the first methodology to allow a network-wide 
unbiased comparison of community structure across multiple realizations of a 
network. Scaled inclusivity differs in several ways from other approaches for 
evaluating the consistency of community structure. In scaled 
inclusivity, the partitions of multiple networks are compared to a reference 
partition with no a priori assumptions about which communities match one 
another. Every node is independently assigned a scaled value for each 
comparison to the reference partition based on how well the communities 
containing that node match in the two partitions. It is thus possible to 
quantify the consistency of a node's classification across the partitions of 
several networks. One advantage of this approach is the inclusion of all nodes 
in the final analysis. Instead of only indicating the most consistent 
communities, scaled inclusivity assigns a consistency value to every node in 
the network. In addition, because the method makes no assumptions about the 
relationship among the networks, scaled inclusivity is equally applicable to 
cross-sectional (across-network) and longitudinal (across time) analyses. 
Finally, scaled inclusivity can be applied to datasets of any size.

We tested our algorithm on one simulated dataset and one real dataset. The 
simulated data used randomly generated communities which were manually assigned 
to the most populous cities in the United States to yield geographically 
contiguous communities. We found that the areas with the highest scaled 
inclusivity values (and thus the highest consistency of classification) were 
Texas, Florida, the Northeast, Southern California, and Northern California. 
Florida and the Northeast are highly consistent because of the geographic 
constraints placed on the community definitions when generating the networks. 
The most natural way to put these regions into 
geographical communities is to start from the edges of the map and work inward. 
For both California regions, the density of nodes determined the consistency 
of classification. The generally limited community size and the concentration 
of cities in Southern California caused those cities to be placed in 
the same community in nearly every map. A similar effect can be seen 
in Northern California as well. While there is not such a simple explanation 
for the high scaled inclusivity in Texas, the central region of Texas that has 
the highest consistency is grouped together in every partition, as can be seen 
in the Appendix. In addition, the Midwest, Southeast, and West varied more in 
their community boundaries due to changing community size; this is reflected in 
their lower scaled inclusivity values. Overall, the data seem to reflect the 
consistency of the partitions.

We also compared college football networks for the years 1995 to 2009. The 
scaled inclusivity for the true conference alignment closely follows what would 
be expected. The SEC, PAC-10, and Big Ten teams all had the highest possible 
scaled inclusivity values, which is to be expected because these conferences 
did not change membership during the time period studied. The teams in the Big 
12 had only slightly lower values; the additions to the Big 8 in 1996 made a 
small but noticeable impact on the consistency of the conference as a whole. 
Conferences that have 
changed drastically since 1995 have much lower values, including the Big East, 
the WAC, Conference USA, and the Sun Belt. As with the simulated data, the 
scaled inclusivity values for the college football network fit with the known 
changes in community structure.

The results were less promising for the analysis of the football network 
performed on the QCut results. 
Due to the misclassifications of some teams (especially independents) and the 
combinations of conferences in certain years, the most consistent conferences 
are not so easily identified. Some of these problems are probably not a result 
of the community structure algorithm but rather of the network itself. 
Independent teams frequently play many teams from conferences located nearby, 
and conferences with overlapping territory also form rivalries with many 
inter-conference games. As a result, the inclusion of independent teams and 
the combination of multiple conferences is to be expected. The scaled 
inclusivity values are still generally lower for the less consistent 
conferences, but the differences are less clear. 
The results of this analysis are more difficult to interpret 
than similar results from the true conference alignments.

Among the difficulties inherent in identifying consistency or change in network 
community structure is the limitation of current algorithms. Because finding a 
network partition with optimal modularity has been shown to be an NP-complete 
problem, any algorithm based on modularity that runs 
in a reasonable amount of time will yield imperfect results. In many cases, 
the solution given may vary across multiple runs on the same network. In 
addition, there is evidence to suggest that the problem of optimizing Q has a 
degenerate search space, meaning many solutions exist with near-optimal values 
of Q that have significantly different community structure \cite{degenerate}. 
To mitigate these effects on our analysis, we ran the modularity algorithm 
(QCut) multiple times for each network and selected the best results. 
This prevents a single improbable run from skewing the results of the analysis. 
However, we have seen that imperfections in the reference partition can unduly 
influence the final scaled inclusivity values in situations where a node is 
consistently classified in all other networks. To counter this problem, we 
generated a weighted average of the scaled inclusivity with each partition as 
the referent partition, a map that uses negative values to evaluate the 
consistency of nodes not included in the referent's community, and a map 
showing which referent partition yielded the highest scaled inclusivity value 
for each node independently. All of these methods minimize the bias from a 
single partition on the final evaluation of consistency. This was demonstrated 
in greater detail with the community containing Texas in Section IV. 

In this paper, we propose a method for comparing the consistency of community 
structure across different realizations of a network, be they the same network 
over time or multiple realizations of a single network. In particular, we 
propose a method to describe how consistently each node is part of the same 
community across different partitions using a metric called scaled inclusivity. 
This method enables us to identify which nodes tend to remain in the same 
community in different partitions, forming a "core" of that community. 
Likewise, the method also allows identification of nodes that become part of 
different communities across partitions.

Community structure analysis is a growing field, and new algorithms to find the 
optimal partition of a network are being developed. Improved algorithms promise 
better partitioning of complex networks, which would render our method more 
effective in analyzing the consistency of the true community structure across 
networks. In addition, some recent community structure algorithms permit a node 
to be a member of multiple communities \cite{Palla2, Baumes}. This is thought 
to better characterize the community structure of some real-world networks. 
Basing our analysis on such algorithms is an avenue for potential future 
research. Finally, a key future application of the scaled inclusivity algorithm 
is its use on biological networks. The study of networks across subjects in 
research studies is rapidly 
growing, and the ability to compare the community structures of different 
populations promises to be a valuable research tool. Some of the growing 
subfields of biological network science include complex brain networks, protein 
networks, genome networks, and metabolic networks.

\begin{acknowledgments}
This work was supported in part by the National Institute of Neurological 
Disorders and Stroke (NS070917 and NS039426-09S1) and the Translational Science 
Institute of Wake Forest University (TSI-K12).
\end{acknowledgments}


\appendix*
\section{Community maps of reference community containing Texas for all other 
simulated networks}
\begin{figure*}[]
\begin{center}
\subfigure[Map 1]{\label{fig:edge-a}
\includegraphics[width=35mm,height=20mm]{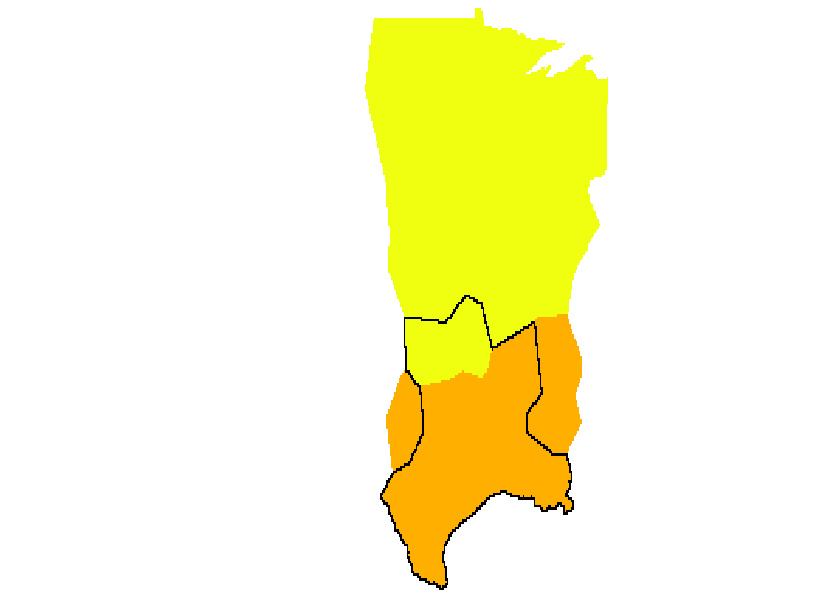}}
\subfigure[Map 2]{\label{fig:edge-b}
\includegraphics[width=35mm,height=20mm]{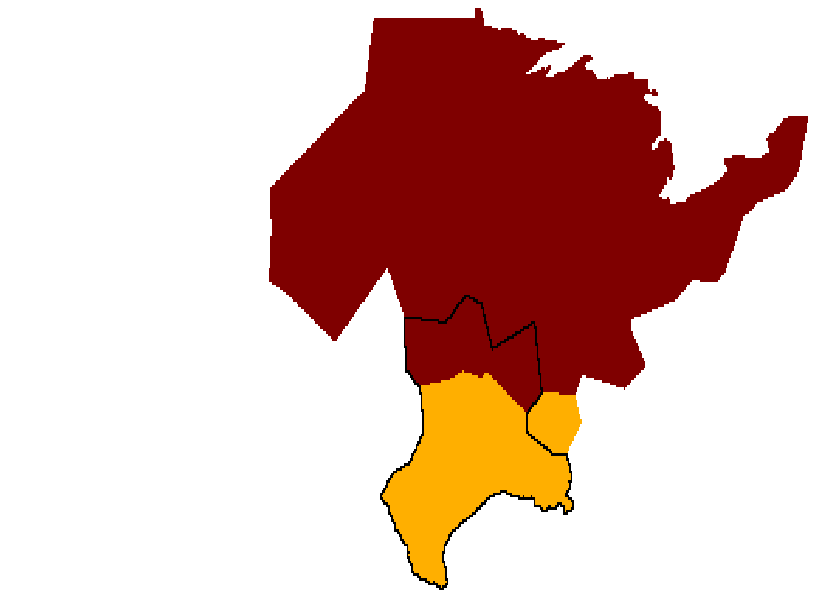}}
\subfigure[Map 3]{\label{fig:edge-c}
\includegraphics[width=35mm,height=20mm]{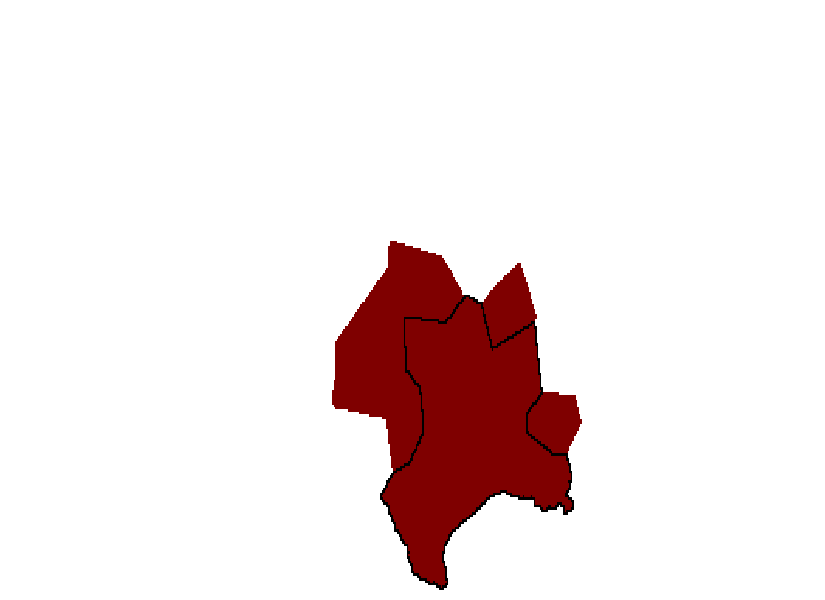}}
\subfigure[Map 4]{\label{fig:edge-d}
\includegraphics[width=35mm,height=20mm]{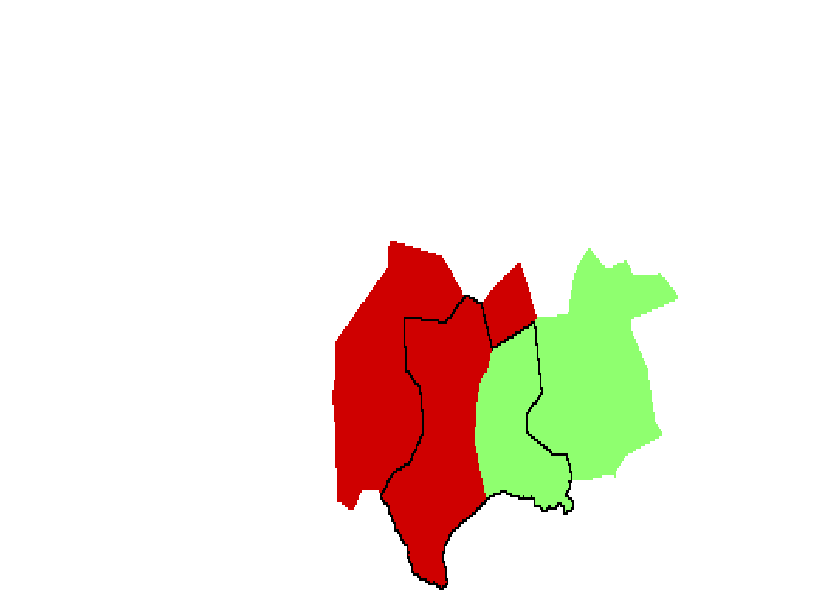}}
\subfigure[Map 5]{\label{fig:edge-e}
\includegraphics[width=35mm,height=20mm]{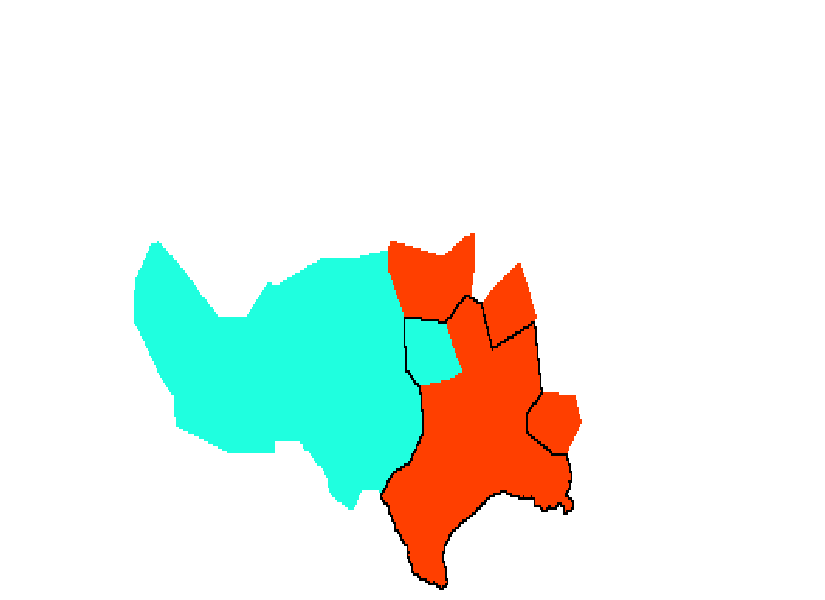}}
\subfigure[Map 6]{\label{fig:edge-f}
\includegraphics[width=35mm,height=20mm]{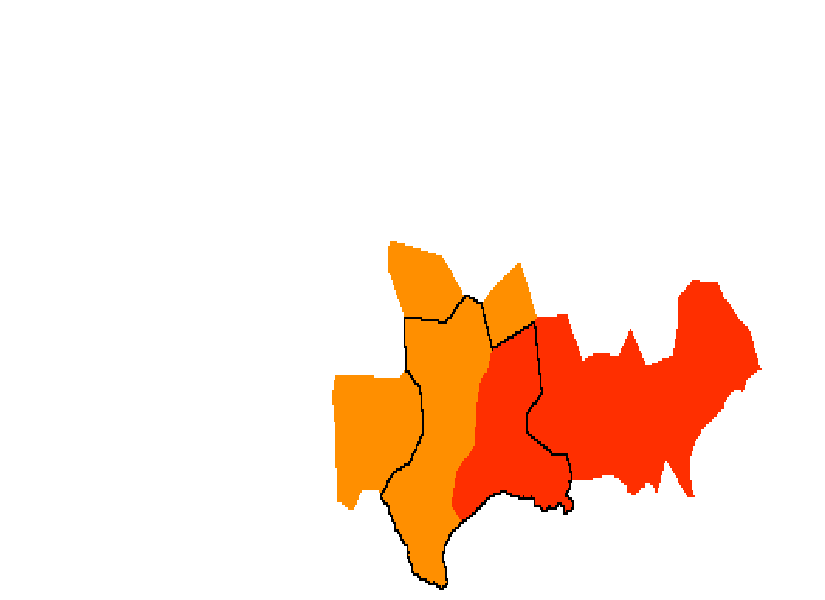}}
\subfigure[Map 7]{\label{fig:edge-g}
\includegraphics[width=35mm,height=20mm]{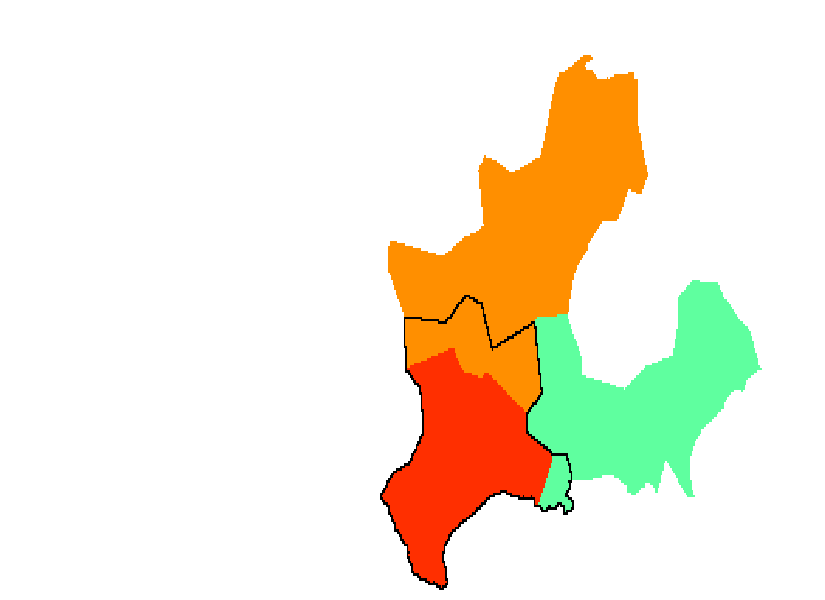}}
\subfigure[Map 8]{\label{fig:edge-h}
\includegraphics[width=35mm,height=20mm]{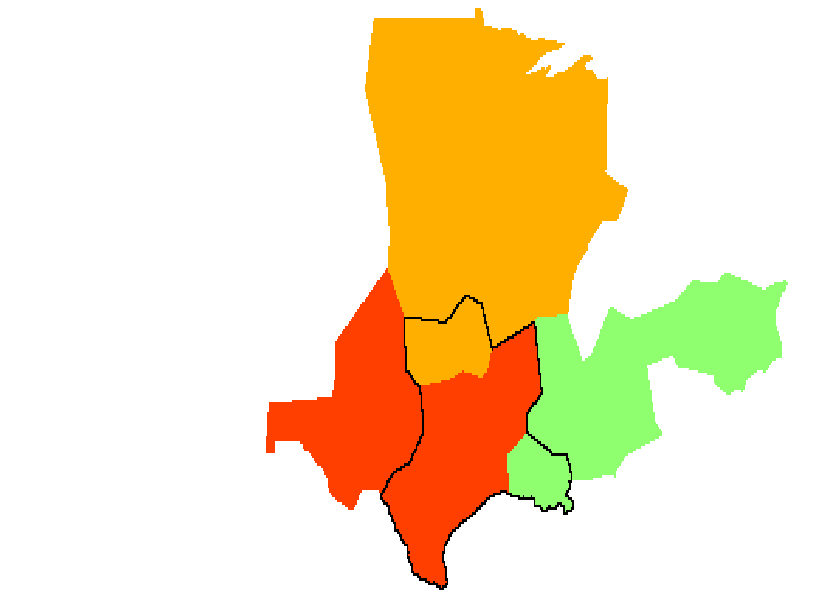}}
\subfigure[Map 9]{\label{fig:edge-i}
\includegraphics[width=35mm,height=20mm]{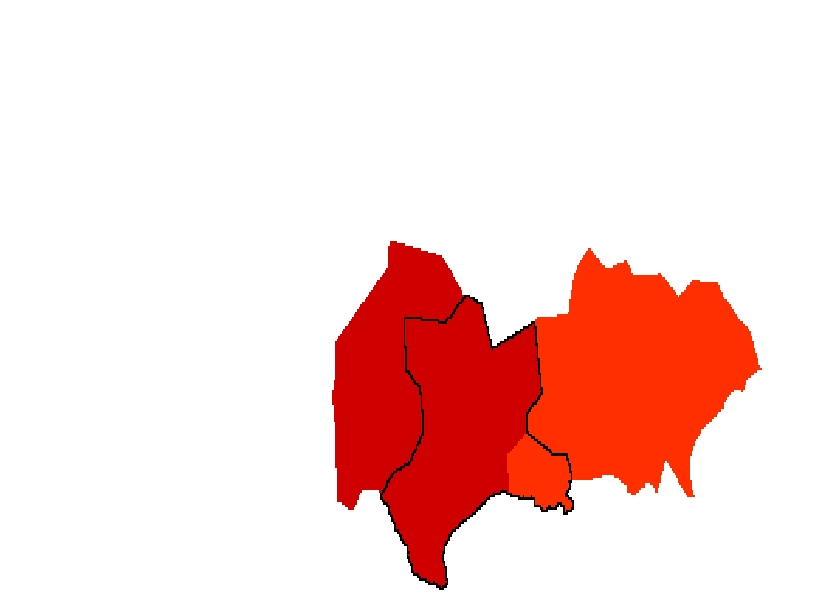}}
\subfigure[Map 10]{\label{fig:edge-j}
\includegraphics[width=35mm,height=20mm]{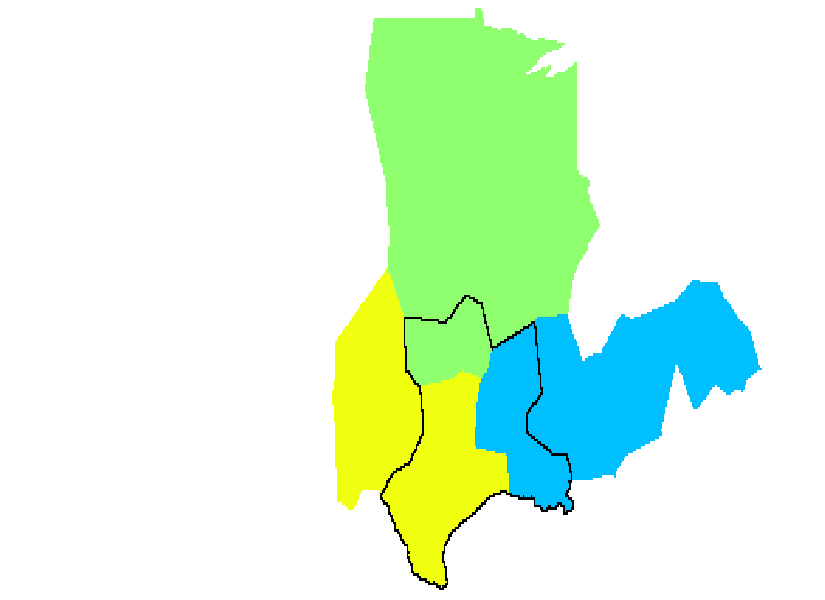}}
\subfigure[Map 11]{\label{fig:edge-k}
\includegraphics[width=35mm,height=20mm]{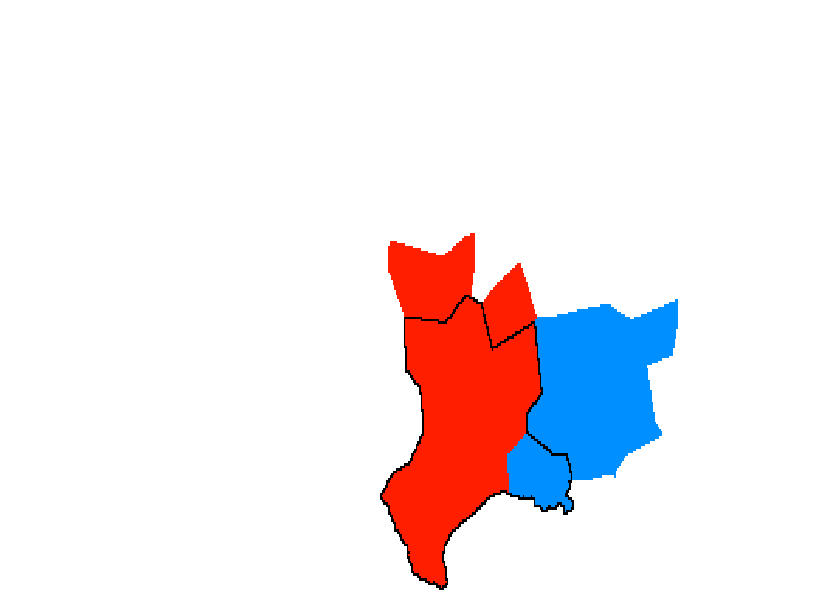}}
\subfigure[Map 12]{\label{fig:edge-l}
\includegraphics[width=35mm,height=20mm]{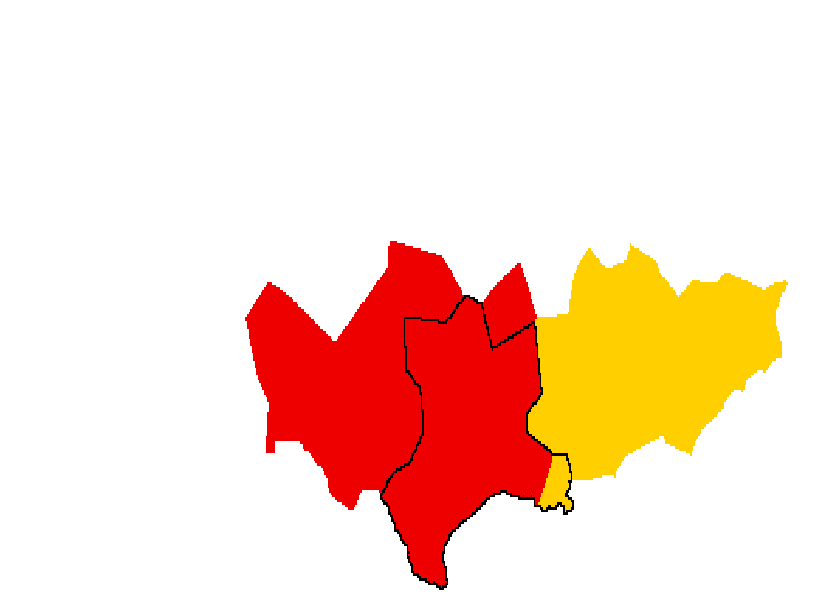}}
\subfigure[Map 13]{\label{fig:edge-m}
\includegraphics[width=35mm,height=20mm]{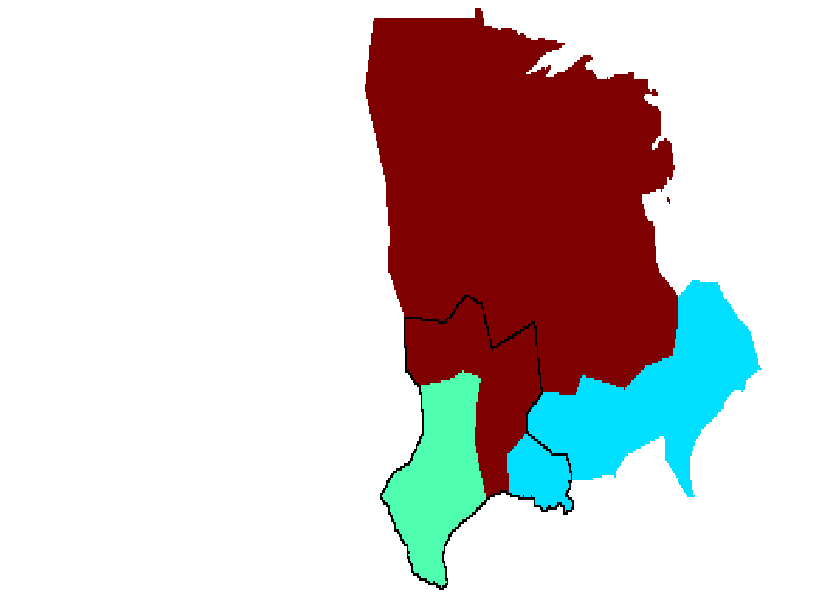}}
\subfigure[Map 14]{\label{fig:edge-n}
\includegraphics[width=35mm,height=20mm]{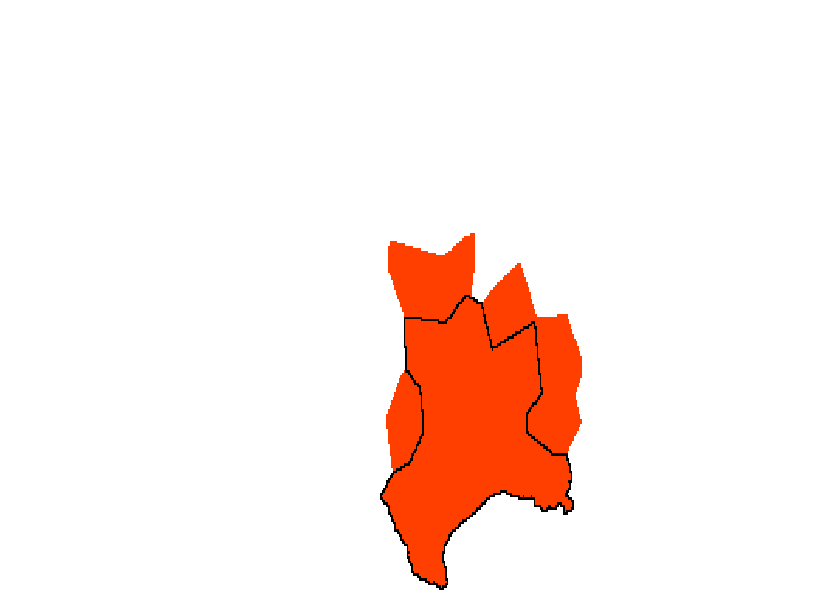}}
\subfigure[Map 15]{\label{fig:edge-o}
\includegraphics[width=35mm,height=20mm]{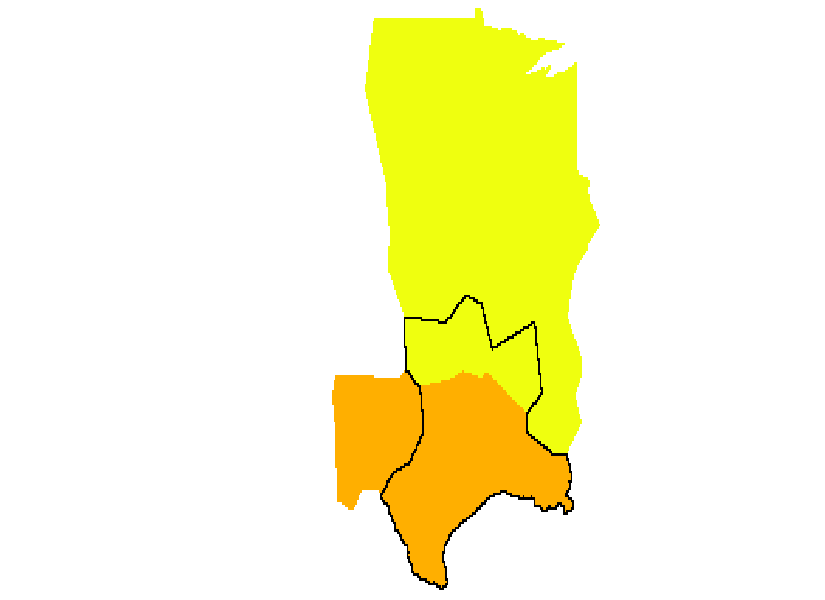}}
\subfigure[Map 16]{\label{fig:edge-p}
\includegraphics[width=35mm,height=20mm]{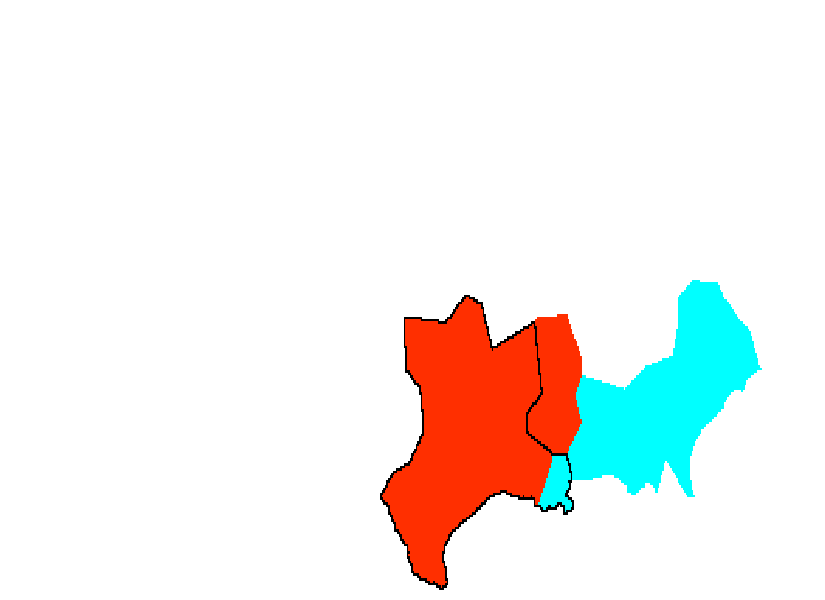}}
\subfigure[Map 17]{\label{fig:edge-q}
\includegraphics[width=35mm,height=20mm]{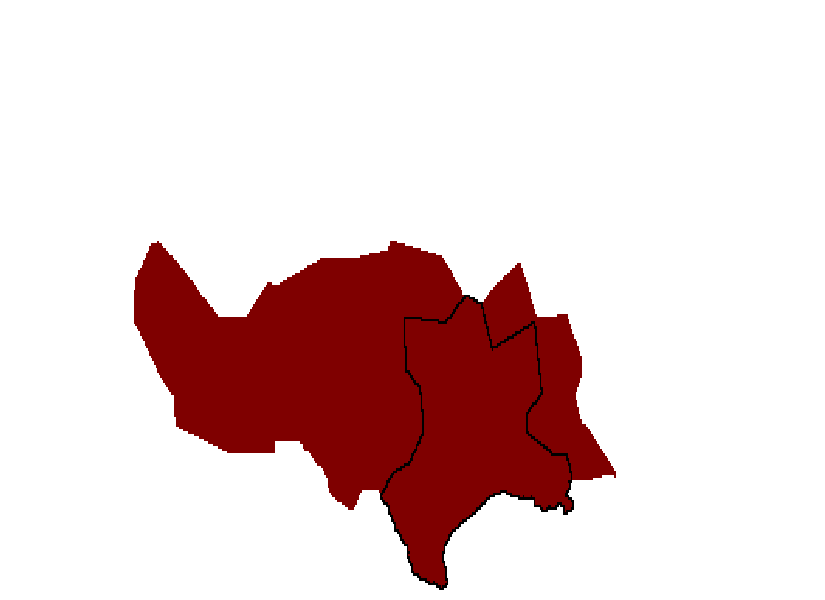}}
\subfigure[Map 18]{\label{fig:edge-r}
\includegraphics[width=35mm,height=20mm]{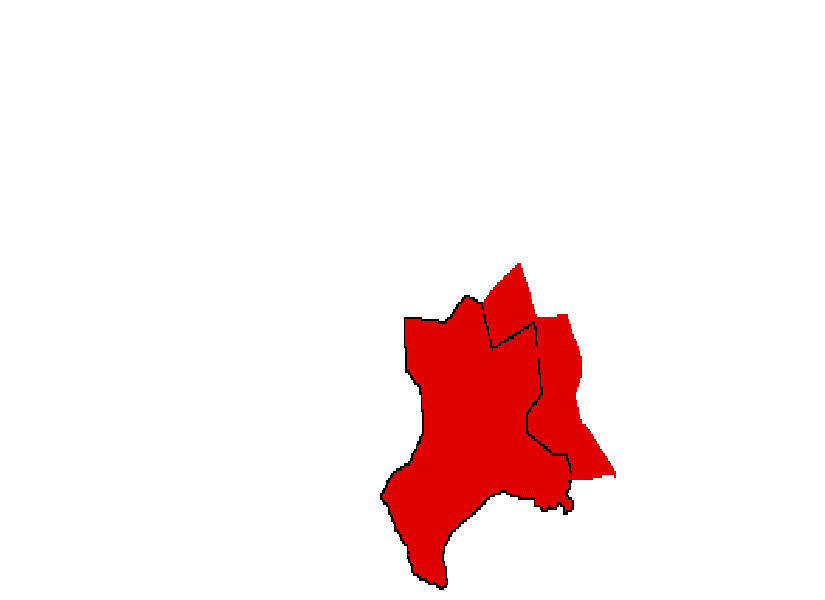}}
\subfigure[Map 19]{\label{fig:edge-s}
\includegraphics[width=35mm,height=20mm]{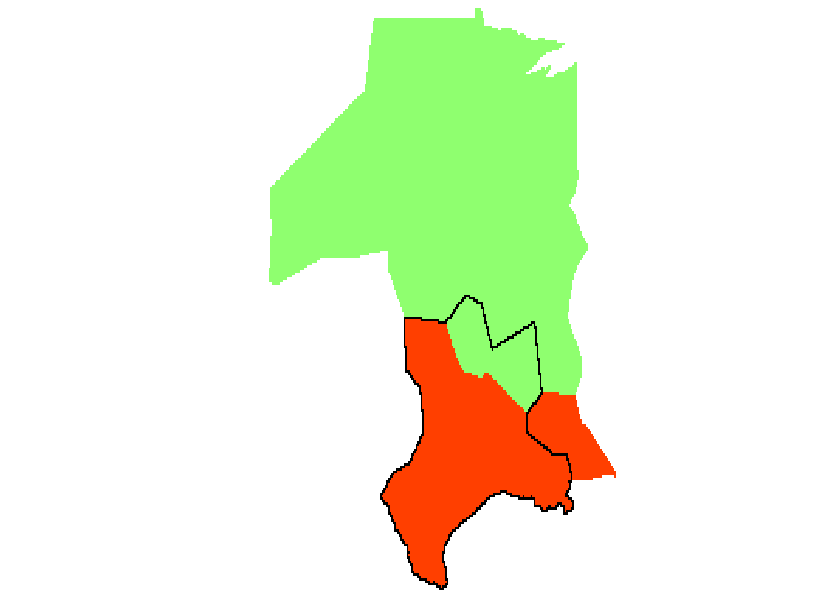}}
\subfigure[Map 20]{\label{fig:edge-t}
\includegraphics[width=35mm,height=20mm]{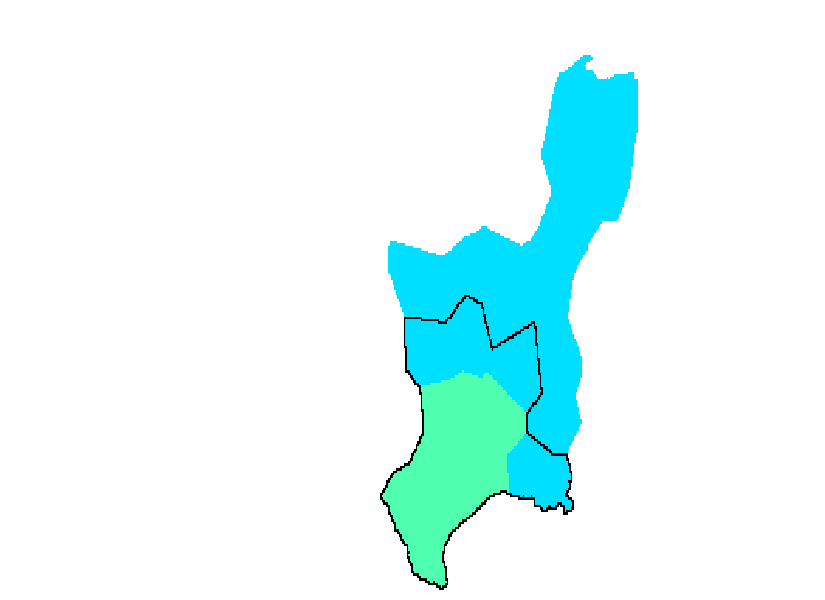}}
\subfigure[Map 21]{\label{fig:edge-u}
\includegraphics[width=35mm,height=20mm]{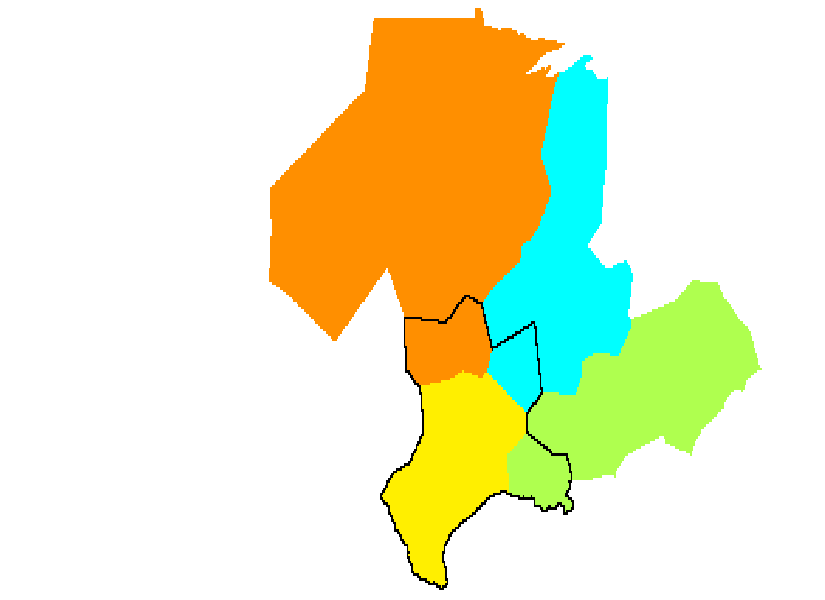}}
\subfigure[Map 22]{\label{fig:edge-v}
\includegraphics[width=35mm,height=20mm]{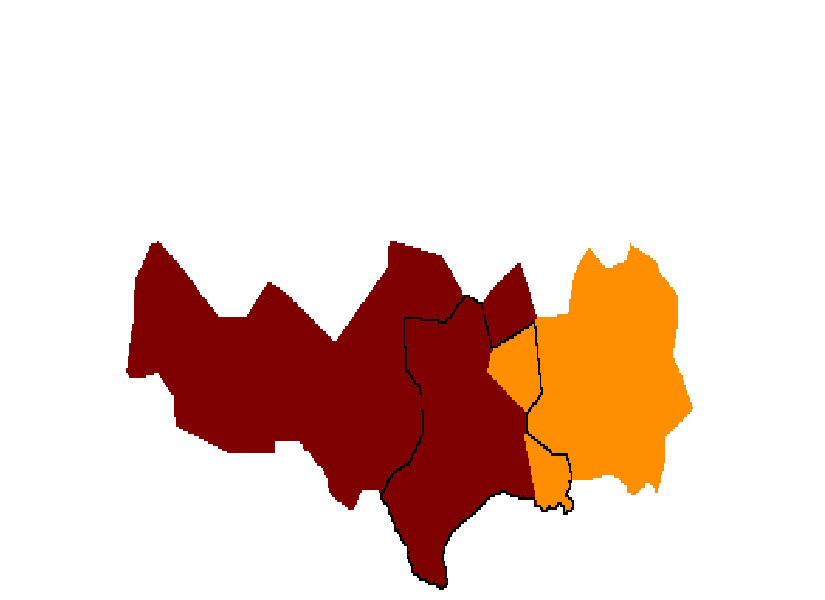}}
\subfigure[Map 23]{\label{fig:edge-w}
\includegraphics[width=35mm,height=20mm]{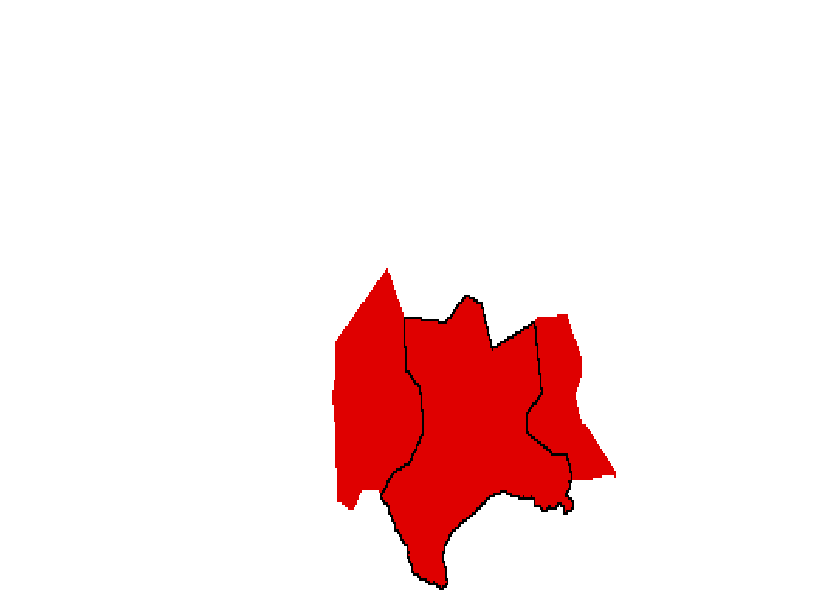}}
\subfigure[Map 24]{\label{fig:edge-x}
\includegraphics[width=35mm,height=20mm]{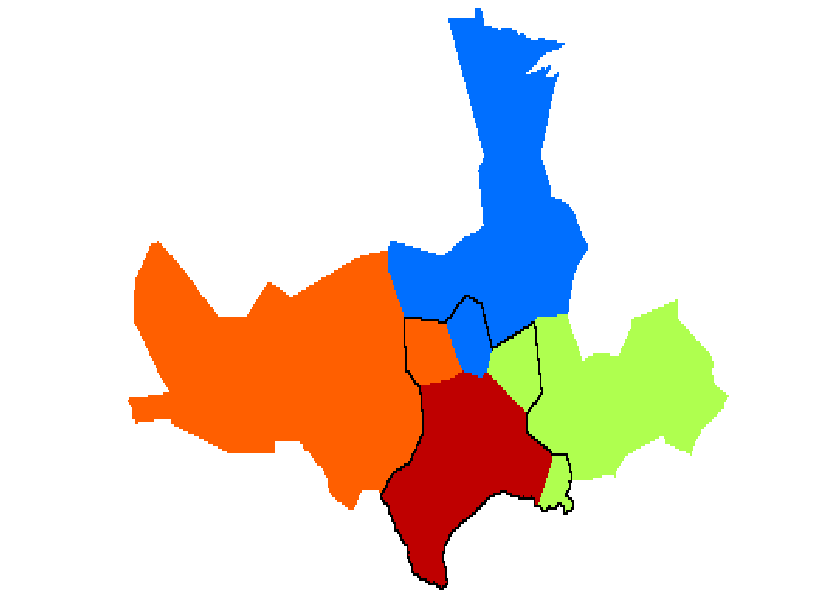}}
\subfigure[Map 25]{\label{fig:edge-y}
\includegraphics[width=55mm,height=35mm]{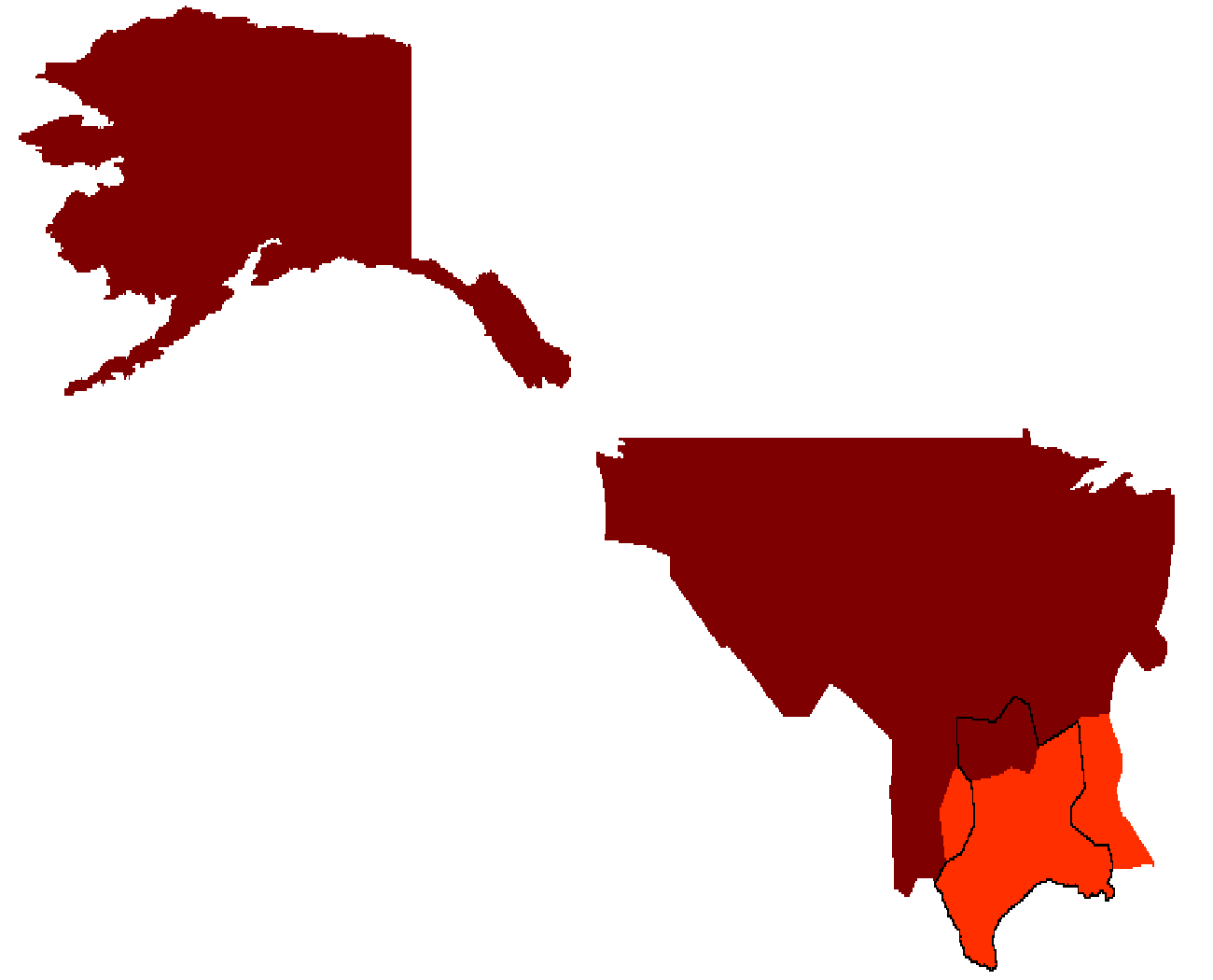}}
\subfigure[Map 27]{\label{fig:edge-z}
\includegraphics[width=35mm,height=20mm]{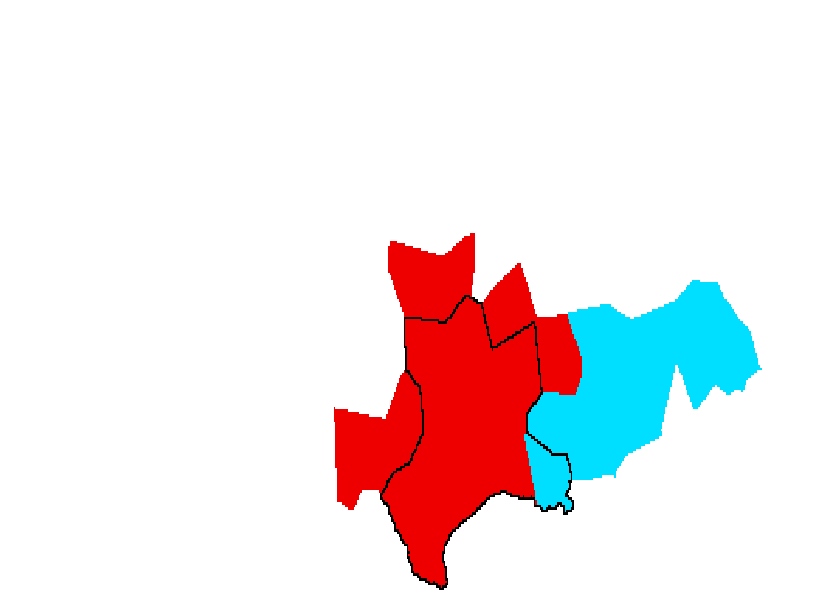}}
\subfigure[Map 28]{\label{fig:edge-aa}
\includegraphics[width=35mm,height=20mm]{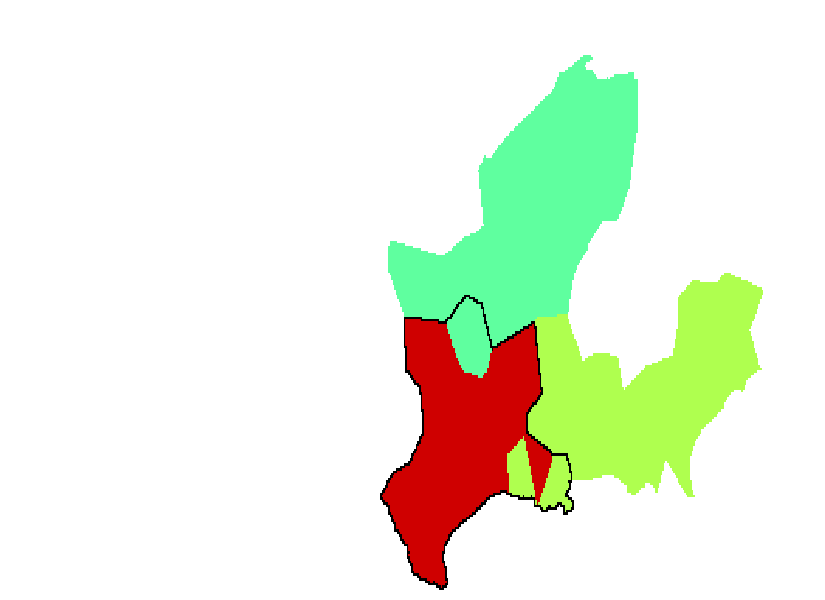}}
\subfigure[Map 29]{\label{fig:edge-bb}
\includegraphics[width=35mm,height=20mm]{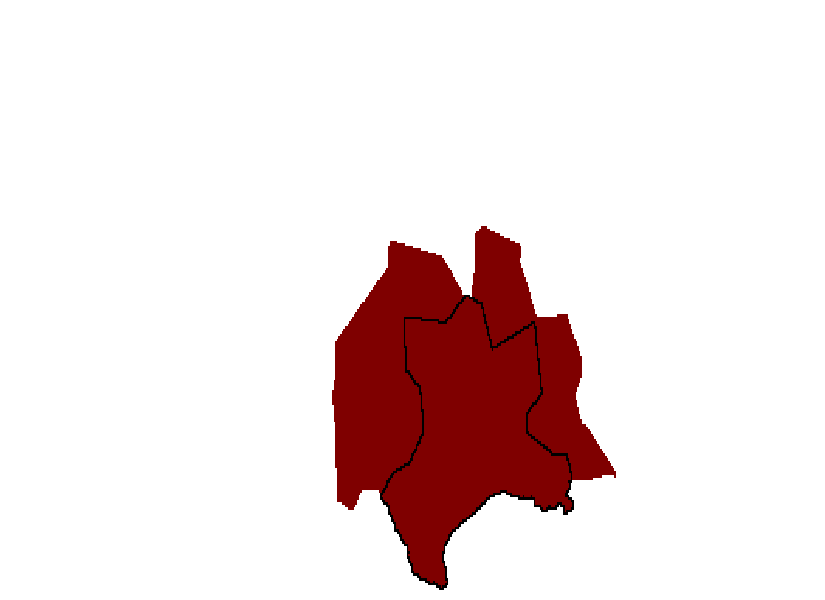}}
\subfigure[Map 30]{\label{fig:edge-cc}
\includegraphics[width=35mm,height=20mm]{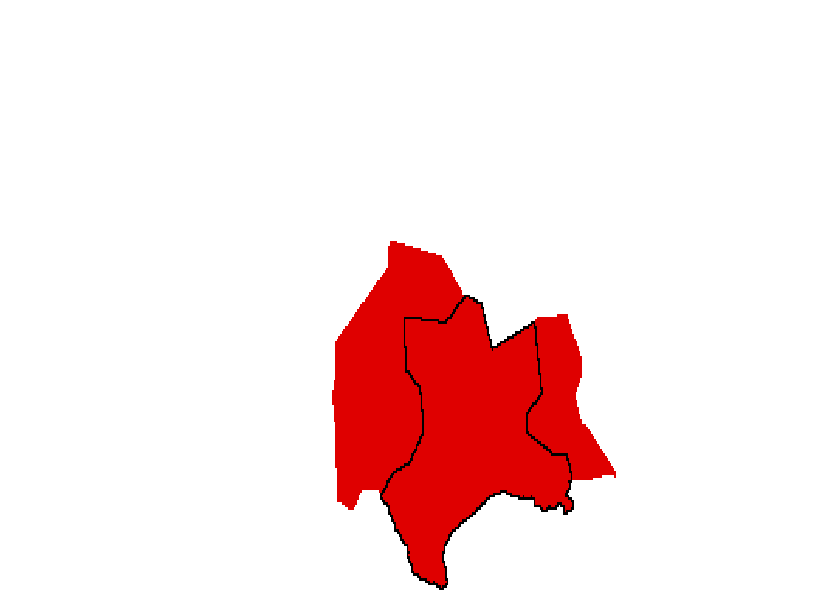}} \\
\end{center}
\caption{(Color online) This figure shows the simulated communities for each 
network that overlap with the community containing Texas in the reference 
community (network 26). That community is outlined in each image. The colors of 
the other communities in each image are arbitrarily assigned. 
\label{Texas_all_partitions}}
\end{figure*}

\clearpage


%

\end{document}